# Redesigning Large-Scale Multimodal Transit Networks with Shared Autonomous Mobility Services

Max T.M. Ng[a], Hani S. Mahmassani[a,*],
Ömer Verbas[b], Taner Cokyasar[b], Roman Engelhardt[c]

[a] *Department of Civil and Environmental Engineering,
Transportation Center, Northwestern University
600 Foster Street, Evanston, IL 60208, United States*

[b] *Argonne National Laboratory,
9700 Cass Avenue, Lemont, IL 60439, United States*

[c] *Chair for Traffic Engineering and Control,
Technical University of Munich, 80333 Munich, Germany*

## Abstract

This study addresses a large-scale multimodal transit network design problem, with Shared Autonomous Mobility Services (SAMS) as both transit feeders and an origin-to-destination mode. The framework captures spatial demand and modal characteristics, considers intermodal transfers and express services, determines transit infrastructure investment and path flows, and generates transit routes. A system-optimal multimodal transit network is designed with minimum total door-to-door generalized costs of users and operators, satisfying transit origin-destination demand within a pre-set infrastructure budget. Firstly, the geography, demand, and modes in each zone are characterized with continuous approximation. The decisions of network link investment and multimodal path flows in zonal connection optimization are formulated as a minimum-cost multi-commodity network flow (MCNF) problem and solved efficiently with a mixed-integer linear programming (MILP) solver. Subsequently, the route generation problem is solved by expanding the MCNF formulation to minimize intramodal transfers. The model is illustrated through a set of experiments with the Chicago network comprised of 50 zones and seven modes, under three scenarios. The computational results present savings in traveler journey time and operator cost demonstrating the potential benefits of collaboration between multimodal transit systems and SAMS.

---

* Corresponding author. E-mail address: masmah@northwestern.edu.

# 1 Introduction

## 1.1 Motivation

Public transit has traditionally served as the primary mobility solution in cities, supported by its economies of density. However, transit usage has been declining, especially in North America. This decline is attributable to long-term structural trends, including changing land use patterns and increased competition from on-demand private mobility platforms. Recent trends associated with the COVID-19 pandemic, such as reduced services and the rise of spatially and temporally flexible hybrid/remote work arrangements (Tahlyan et al., 2022), have only exacerbated this issue (Kar et al., 2022), increasing dependence on the private automobile despite its higher costs for users (fuel and maintenance), society (delay due to congestion), and the environment (15% of global carbon emissions contributed by road transport (Ritchie, 2020)). Autonomous vehicles portend a future of Shared Autonomous Mobility Services (SAMS) with convenient on-demand point-to-point mobility. However, this convenience is predicted to induce demand and, thus, more traffic and emissions, unless transit options are also improved or reimagined to leverage SAMS and other emerging technologies (Levin and Boyles, 2015; Xu et al., 2019).

A natural yet underdeveloped solution is to redesign transit systems to leverage the service and cost flexibility of SAMS. Despite direct competition with the transit system for certain trips, shared autonomous vehicles (SAVs) can provide first-mile, last-mile access to connect people to transit systems. Conventional transit modes such as rail, bus rapid transit (BRT), and bus lines should then be reconfigured to serve the new demand patterns, for example, by optimizing their frequencies and alignments with respect to travel times, waiting times, and operating costs. This would expand the transit catchment areas and enhance ridership, strengthening its economies of density and justifying higher service levels without excessive investment in the first and last-mile service. Recognizing this complementarity in the design process would leverage the respective strengths of both types of service in meeting a wide range of mobility needs. Building on the transit-SAMS joint design problem proposed by Pinto et al. (2020), this study presents an efficient framework to comprehensively design a multimodal transit network that incorporates SAMS as both feeder and direct services.

## 1.2 Problem description

This paper addresses a large-scale multimodal transit network design problem with SAMS, that includes ride-pooling SAVs functioning as both transit feeders and an origin-to-destination mode. It aims to minimize the total generalized door-to-door costs, which include user costs (access, waiting, travel, transfer) and operator costs (operation and emissions), governed by constraints to satisfy origin-destination (O-D) demand within a predefined infrastructure budget. This is equivalent to maximizing social welfare.

There are several layers of complexity to this problem. Firstly, the transit network design problem (TNDP) is NP-hard (Magnanti and Wong, 1984), resulting in a solution space that expands exponentially with the number of network nodes. Furthermore, potential non-linearity arises from the interdependency between path-level and link-level costs, while the problem is in general non-convex (Newell, 1979). Lastly, multimodality introduces interactions across modes, in terms of competition for demand and infrastructure budget, and spaces, for economies of scale. Changes in one mode's service can trigger changes in demand for other modes; for example, consolidating and reducing commuter rail routes could shift peripheral demand towards origin-to-destination SAMS. These challenges preclude exact solution methods in favor of heuristic approaches that allow consideration of essential features of the problem in representing modal interactions inherent to the large scale of the problem. In light of these difficulties, a computationally efficient solution method becomes crucial for application to city-scale networks.

This intricate problem can be decomposed into three interconnected sub-problems:
1. **High-level multimodal network modeling**: constructing a zonal network model that captures the spatial distribution of demand, as well as the service class, cost structures, and existing networks of each mode. The optimal mode to serve the demand is then determined. High-density corridors like rail and BRT require right-of-way investment but offer faster service and lower average cost. In contrast, SAMS for feeder or origin-to-destination services offer higher flexibility at a higher average cost. The mid-tier bus services utilize road infrastructure but run more slowly. Furthermore, the model should differentiate local and express services.
2. **Optimal link investment and path flows**: determining the system-optimal infrastructure investment combination across modes within a capital budget and multimodal paths for each O-D pair. This sub-problem also accommodates competition and transfers across modes.
3. **Route design**: designing the routes of each transit mode to minimize intramodal transfers, thereby providing most passengers with seamless one-seat rides.

## 1.3 Contributions

This study introduces a high-level multimodal transit network design and route generation framework. This framework captures the interaction among demand, geospatial, and modal characteristics to develop a system-optimal multimodal transit network and generate routes. These routes then serve as the foundation for the subsequent joint transit-SAMS service level optimization (frequency and SAV fleet size) and simulation (SAV fleet and traveler assignment) problem (Pinto et al., 2020).

This study presents conceptual, theoretical, and methodological contributions. First, it models and solves a large-scale transit network redesign problem that simultaneously considers *multiple* transit modes and SAMS while optimally generating new infrastructure links. The formulation allows planners to retain certain (or no) elements of existing infrastructure and services while redesigning others. The formulation corresponds to a new conceptualization of how emerging mobility services (SAMS) could interact with and complement various forms of conventional and hybrid transit modes. As such, it represents a timely contribution to the existing literature for transit planning in conjunction with emerging mobility services. Second, the zonal connection optimization problem is formulated as a minimum-cost multi-commodity network flow (MCNF) problem, where the commodities correspond to origins for efficient computation of system-optimal solutions fulfilling all demand. We also consider connection candidates strategically to control the solution space and represent local/express services. The formulation efficiency is demonstrated with a large-scale network in the Chicago metropolitan area in the U.S., in which the mixed integer linear programming (MILP) problems are solved within one to two days. The third contribution is the formulation of the route generation problem as an expanded MCNF that leverages fixed path flows–obtained from the zonal optimization model–to minimize intramodal transfers globally. The advantage of this method is benchmarked against myopic heuristics also developed by the authors.

Figure 1 summarizes the components of the methodology presented in this paper for the high-level redesign problem. and its interaction with the lower-level joint transit-SAMS service level optimization-simulation problem. Two key aspects define the scope of the problem addressed in this paper. First, only *interzonal* trips are modeled and evaluated in this study, as both interzonal and intrazonal trips would be considered in the next-level optimization-simulation problem. Second, at a strategic level, the model considers a steady state of demand and transport operations, leaving the more detailed time-varying elements to the next-level optimization-simulation results.

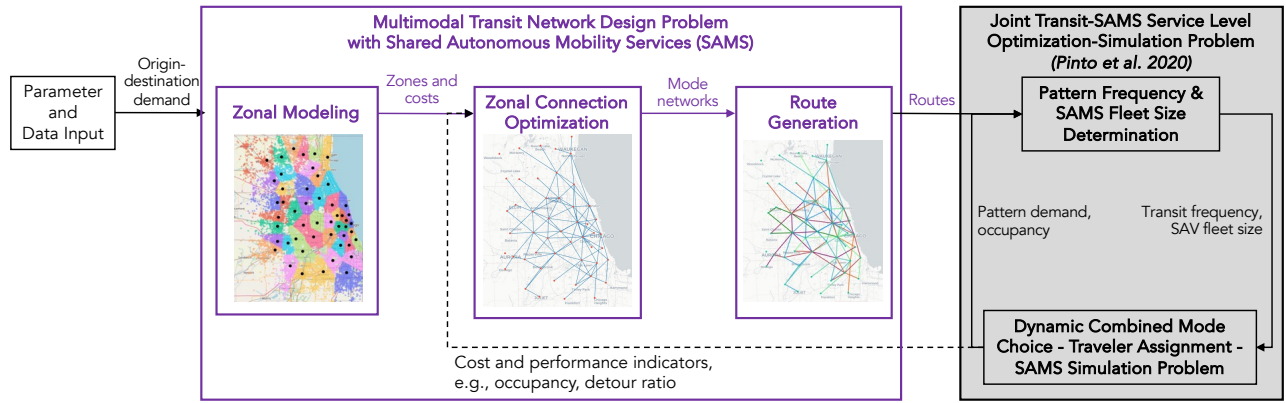

Figure 1. Multimodal transit network design problem with shared autonomous mobility services (SAMS) and its interactions with the joint transit-SAMS service level optimization-simulation problem (Pinto et al., 2020).

The remainder of the paper is structured as follows. Section 2 presents an overview of the modeling framework and reviews relevant literature. Section 3 introduces the mathematical formulation of the network models for zonal connection optimization and route generation. The Nomenclature in Section 3.1 lists the mathematical notation used in the paper. Section 4 details the experimental design using real-world data in the Chicago metropolitan area, while Section 5 showcases the computational results and case studies. Section 6 concludes the paper.

## 2 Background

### 2.1 Transit network design problem

The transit network design problem (TNDP) is a strategic planning problem involving the design of transit routes based on metrics relevant to travelers, operators, or both, often to minimize total generalized costs (Newell, 1979). Research in the last few decades resulted in a rich literature (including seminal works: Baaj and Mahmassani, 1995; Ceder and Wilson, 1986; Clarens and Hurdle, 1975).

Recent efforts in designing modern transit networks include continuous approximation and discrete models. Continuous approximation models assume idealized networks (such as radial and grid) with general parameter distribution (including demand density), yielding analytical solutions (Daganzo, 2010; Badia et al., 2014; Chen et al., 2015). Discrete models consider discrete spatial and demand parameters. Due to their combinatorial nature, metaheuristics are primarily used and optimize networks of limited sizes (see reviews by Iliopoulou et al. (2019)). Alternatively, it can be formulated as linear programs for exact solutions (Borndörfer et al., 2007; Cancela et al., 2015), in which case multi-commodity network flow models (O-D pairs as commodities) ensure demand satisfaction but introduce a tremendous number of variables. For a detailed discussion of recent TNDP variants and solutions, see the review by Durán-Micco and Vansteenwegen (2022).

This study computes costs in each zone with a continuous approximation approach, and then formulates the TNDP as a multi-commodity minimum cost network flow problem with only origins (instead of O-D pairs) as commodities to reduce the number of variables, leading to an efficient MILP formulation to handle city-scale transit networks.

## 2.2 Multimodal transit network design

A growing literature explores multimodal transit network design, involving interaction across modes, such as rapid transit, bus, and ride-hailing. Cipriani et al. (2012) propose heuristics for large-scale network design with a linear relationship between ridership and speed. Modes are often characterized with continuous approximation (Yineng Wang et al., 2022), and intermodal transfers are modeled with virtual links in multimodal network topology (Fan et al., 2022; Yu Wang et al., 2022), while some works optimize transit hub locations (Ye et al., 2021; Yuan and Yu, 2018). This paper utilizes a general multimodal network representation (Figure 2), where demand originates from and terminates at zonal nodes, and flows through arcs to/from and between modal nodes. Starting, ending, interzonal travel, and intermodal transfer costs are incurred on the arcs.

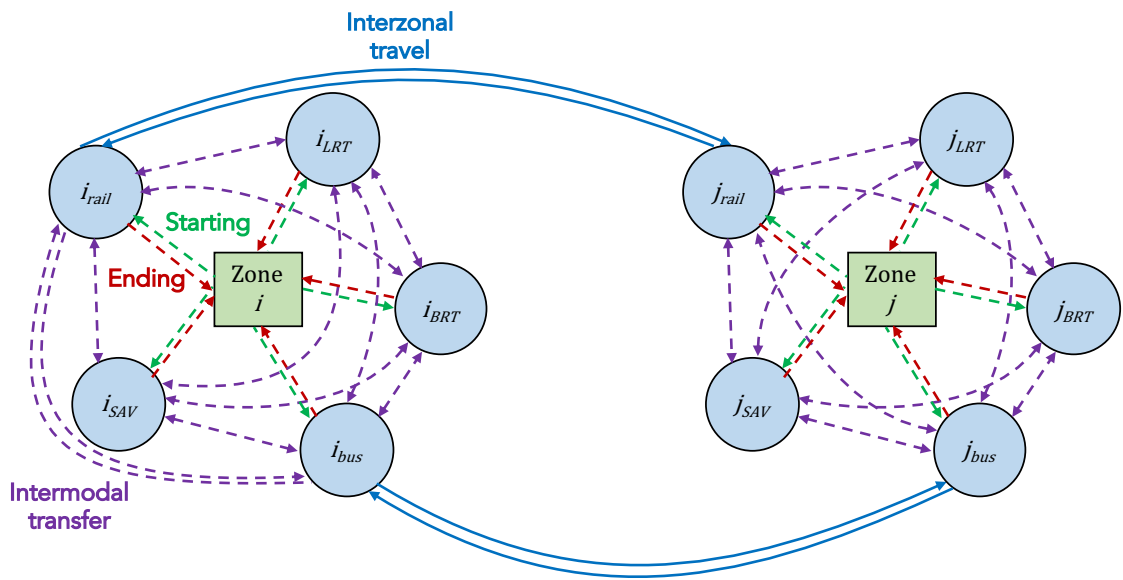

Figure 2. Multimodal transit network representation *(not showing all connections)*.

Fixed guideway transit (including heavy rail, light rail, and BRT) requires dedicated rights-of-way and infrastructure investment. Canca et al. (2019) and Shan et al. (2021) consider the infrastructure costs of link addition in the objective. These binary link-level fixed costs are inconsistent with path-level variable costs, leading to a discrete/mixed network design problem (Luathep et al., 2011; Zhang et al., 2014; see reviews: Farahani et al., 2013). This study handles the infrastructure budget constraint for new links of different modes in the MILP formulation, enabling simultaneous optimization of link addition and flows across modes.

The first-mile, last-mile problem for transit can be mitigated with well-designed feeder services (Sivakumaran et al., 2014). Fixed-route and on-demand feeders offer lower generalized costs depending on the demand and service parameters (Calabrò et al., 2023). Demand-responsive transit is valuable on sparser demand and can complement main-line services (Liu and Ouyang, 2021; Luo and Nie, 2019; Martínez Mori et al., 2023). In this paper, travelers access and egress interzonal transit modes by walking or SAMS feeders, with feeder costs computed via continuous approximation.

## 2.3 Integrating Shared Autonomous Mobility Services and public transit systems

The autonomous driving capability of SAVs leads to lower operating costs and waiting time, as well as more efficient services (Becker et al., 2020; Hörl et al., 2019; Narayanan et al., 2020), while the shared aspect of SAMS further reduces costs and waiting time with increased detours as a trade-off (Alonso-Mora et al., 2017; Hyland and Mahmassani, 2018). Multimodal transit incorporating autonomous taxis or buses expands transit service coverage and enhances its attractiveness (Badia and Jenelius, 2021; Salazar et al., 2020; Shen et al., 2018). The roles of SAMS in the transit system include cooperation (Cortina et al., 2023; Kumar and Khani, 2022; Steiner and Irnich, 2020) and competition (Mo et al., 2021; Ng and Mahmassani, 2023; Sieber et al., 2020).

Previous approaches to the joint design of transit and SAMS typically determine the SAV fleet size (Dandl et al., 2021; Pinto et al., 2020), while designing transit in terms of a limited number of variables in a single mode group, such as transit frequency (Pinto et al., 2020), link addition (Shan et al., 2021), routes and frequency (Kumar and Khani, 2022), transit network parameter and frequencies (Liu and Ouyang, 2021). This paper considers multiple modes and SAMS as feeders and interzonal mode simultaneously, optimizing simultaneously the mode and routing for each O-D pair and infrastructure link addition.

Pinto et al. (2020) introduce the bi-level joint transit-SAMS service level optimization-simulation framework, which integrates supply-side transit and SAMS service level determination problem at the upper level and demand-side mode and route choice problem at the lower level based on Verbas et al. (2015) and Hyland and Mahmassani (2018). The present paper fills the upper-level gap of transit network design and route generation; given the demand and modal performance as input, the methodology in this paper generates a system-optimal transit skeleton network and routes. These in turn provide the basis for subsequent joint transit frequency and SAMS fleet size optimization, and SAV fleet and mode/route choice simulation.

## 2.4 Clustering in transit network design

Clustering has gained popularity in handling city-scale transit systems and demand data, with previous studies focusing on a customized bus design problem to aggregate demand to clustered groups and corridors (Martínez et al., 2015), with simulation studies in London (Viggiano et al., 2018) and Beijing (Gong et al., 2021). Bahbouh et al. (2017) identify corridors by considering demand locations and angles between O-D lines. Cluster results have been used to compute zone-to-zone journey times and design routes. (Heyken Soares, 2021; Heyken Soares et al., 2019) Other clustering applications include dynamic bus system design (Abdelwahed et al., 2023) and hub locations (Boutarfa and Gok, 2023; Huang et al., 2018; Yuan and Yu, 2018). While these studies identify potential high-density transit corridors, they do not consider multiple modes and economies of scale between contiguous links.

This paper employs *k*-means geospatial clustering to divide the study area into *k* clusters, with an aim to minimize the total distances of travelers to travel from zone centroids to origins/destinations. This constructs a high-level zonal network that captures competition, transfers, and synergy, across and within modes, for the next-step zonal connection optimization.

# 3 Model

This section presents the TNDP to determine system-optimal multimodal interzonal connections that minimize the total generalized costs and establish modal routing that minimizes intramodal transfers. The model considers only *interzonal* trips and a steady state of demand and transport operations.

## 3.1 Nomenclature

The notation is listed in Table 1. Sets are denoted as capital scripted characters (e.g., $\mathcal{M}$ and $\mathcal{Z}$), constants as Greek or capital Roman characters (e.g., $\gamma^a$ and $T_m^a$), and indices and variables as small Roman characters (e.g., $m$ and $f_{om}^\alpha$). Superscripts are qualifiers and subscripts are indices.

Table 1. Notation.

| | |
|---|---|
| ***General – sets and indices*** | |
| $\mathcal{A}$ | Set of micro analysis zones (MAZs), indexed by $a \in \mathcal{A}$ |
| $\mathcal{A}_i$ | Subset of MAZs in cluster zone $i \in \mathcal{Z}$; $\mathcal{A}_i \subset \mathcal{A}$ |
| $\mathcal{L}_m$ | Set of potential links of mode $m \in \mathcal{M}$, indexed by tuple of zones, e.g., $(i,j) \in (\mathcal{Z}, \mathcal{Z})$ |
| $\mathcal{L}_m^\lambda$ | Subset of active links of mode $m \in \mathcal{M}$; $\mathcal{L}_m^\lambda \subset \mathcal{L}_m$ |
| $\mathcal{M}$ | Set of modes, indexed by $m \in \mathcal{M}$; $\mathcal{M} = \mathcal{M}^t \cup \{SAMS\}$ |
| $\mathcal{M}^i$ | Subset of transit modes which require link infrastructures, $\mathcal{M}^i \subset \mathcal{M}^t$ |
| $\mathcal{M}^t$ | Subset of transit modes, $\mathcal{M}^t \subset \mathcal{M}$ |
| $\mathcal{R}$ | Set of routes, indexed by $r \in \mathcal{R}$ |
| $\mathcal{R}_m$ | Subset of routes of transit mode $m \in \mathcal{M}^t$; $\mathcal{R}_m \subset \mathcal{R}$ |
| $\mathcal{U}$ | Set of route segments, indexed by $u \in \mathcal{U}$ |
| $\mathcal{U}_{mij}$ | Subset of route segments of mode $m \in \mathcal{M}^t$ from zone $i \in \mathcal{Z}$ to zone $j \in \mathcal{Z}$; $\mathcal{U}_{mij} \subset \mathcal{U}$ |
| $\mathcal{U}_r$ | Subset of route segments of route $r \in \mathcal{R}$; $\mathcal{U}_r \subset \mathcal{U}$ |
| $\mathcal{Z}$ | Set of zones |
| ***General – parameters*** | |
| $D_{ij}^c$ | Side-to-side distance from zone $i \in \mathcal{Z}$ to $j \in \mathcal{Z}$, approximated by the average center-to-center distance between both zones minus center-to-side distances of both zones |
| $D_m^s$ | Design stop spacing of mode $m \in \mathcal{M}^t$ |
| $E_{od}$ | Demand between origin zone $o \in \mathcal{Z}$ and destination zone $d \in \mathcal{Z}$ |
| $F_{SAMS}^{\alpha,max}$ | Maximum number of starting trips of SAMS in a zone per hour |
| $F_{SAMS}^{\beta,max}$ | Maximum number of ending trips of SAMS in a zone per hour |
| $H_m^d$ | Design headway for transit mode $m \in \mathcal{M}^t$ |
| $H_m^{max}$ | Maximum policy headway for transit mode $m \in \mathcal{M}^t$ |
| $H_m^{min}$ | Minimum feasible headway for transit mode $m \in \mathcal{M}^t$ |
| $L_i$ | Average center-to-side distance of zone $i \in \mathcal{Z}$ |
| $M$ | Big-M (a large constant) |
| $N^a$ | Minimum number of adjacent zones linked to each zone |
| $N^d$ | Minimum number of direct zones linked to each zone |
| $N_a^t$ | Number of trips starting/ending at MAZ $a \in \mathcal{A}$ |
| $N_{mij}^\omega$ | Number of existing links from zone $i \in \mathcal{Z}$ to $j \in \mathcal{Z}$ for infrastructure mode $m \in \mathcal{M}^i$ |
| $P_a$ | Coordinates of MAZ $a \in \mathcal{A}$ |
| $R_m^d$ | Design vehicle occupancy of mode $m \in \mathcal{M}$ (passenger/vehicle) |
| $R_m^{df}$ | Design vehicle occupancy of feeders to access mode $m \in \mathcal{M}$ (passenger/vehicle) |
| $R_m^{max}$ | Maximum feasible vehicle occupancy of mode $m \in \mathcal{M}$ (passenger/vehicle) |
| $R_m^{min}$ | Minimum policy vehicle occupancy of mode $m \in \mathcal{M}$ (passenger/vehicle) |
| $S_{SAMS}^t$ | Total SAV fleet size for interzonal SAMS |
| $S_{SAMS}^\lambda$ | Maximum number of SAVs flowing in a link for interzonal SAMS |
| $T_m^a$ | Average walking time of mode $m \in \mathcal{M}$ |
| $T_m^s$ | Dwell time of mode $m \in \mathcal{M}$ |
| $T_m^{sf}$ | Dwell time for feeders accessing mode $m \in \mathcal{M}$ |
| $T_{m_1 m_2}^t$ | Average transfer time from mode $m_1 \in \mathcal{M}$ to $m_2 \in \mathcal{M}$ |
| $T_m^w$ | Average waiting time of mode $m \in \mathcal{M}$ |
| $T_m^{wf}$ | Average waiting time of feeders to access mode $m \in \mathcal{M}$ |
| $V^a$ | Average walking speed of travelers |
| $V_m^r$ | Average vehicle speed of mode $m \in \mathcal{M}$ |
| $V_m^{rf}$ | Average vehicle speed of feeders to access mode $m \in \mathcal{M}$ |
| $\gamma^a$ | Coefficient converting walking time into travel time |
| $\gamma_m^e$ | Cost of carbon emissions of mode $m \in \mathcal{M}$ per vehicle-revenue-kilometer |
| $\gamma_m^{ef}$ | Cost of carbon emissions of feeders to access mode $m \in \mathcal{M}$ per vehicle-revenue-kilometer |
| $\gamma_m^i$ | Infrastructure cost of mode $m \in \mathcal{M}^i$ per kilometer |
| $\gamma^r$ | Value of travel time per hour |
| $\gamma_m^o$ | Operating cost of mode $m \in \mathcal{M}$ per vehicle-revenue-hour |
| $\gamma_m^{of}$ | Operating cost of feeders to access mode $m \in \mathcal{M}$ per vehicle-revenue-hour |
| $\gamma^w$ | Coefficient converting waiting time into travel time |
| $\phi_m$ | Detour factor for mode $m \in \mathcal{M}$, as the ratio of average distance traveled relative to Euclidean distance |
| $\phi_m^f$ | Detour factor for feeders to access mode $m \in \mathcal{M}$, as the ratio of average distance traveled relative to Euclidean distance |
| $\rho_{SAMS}$ | Design utilization rate of SAVs for interzonal SAMS |
| $\Gamma$ | Infrastructure investment budget |

*Zonal modeling – variables/functions*

| | |
|---|---|
| $c_{mi}^\alpha$ | Starting travel cost to mode $m \in \mathcal{M}$ in zone $i \in \mathcal{Z}$ (in multiples of travel time) |
| $c_{mi}^\beta$ | Ending travel cost from mode $m \in \mathcal{M}$ in zone $i \in \mathcal{Z}$ (in multiples of travel time) |
| $c_{mij}^\lambda$ | Interzonal travel cost of mode $m \in \mathcal{M}$ from zone $i \in \mathcal{Z}$ to $j \in \mathcal{Z}$ (in multiples of travel time) |
| $c_{m_1 m_2 i}^\chi$ | Intermodal transfer cost from mode $m_1 \in \mathcal{M}$ to mode $m_2 \in \mathcal{M}$ in zone $i \in \mathcal{Z}$ (in multiples of travel time) |
| $d_{mi}^f$ | Euclidean distance for feeders to access mode $m \in \mathcal{M}$ in zone $i \in \mathcal{Z}$ |
| $d_{mij}^\lambda$ | Interzonal distance from zone $i \in \mathcal{Z}$ to $j \in \mathcal{Z}$ for mode $m \in \mathcal{M}$ |
| $p_{mi}^f$ | Proportion of riders to access mode $m \in \mathcal{M}$ in zone $i \in \mathcal{Z}$ by feeders |
| $t_{mij}^c$ | Travel time of mode $m \in \mathcal{M}$ through the side-to-side distance from zone $i \in \mathcal{Z}$ to $j \in \mathcal{Z}$ |
| $t_m^d(\cdot)$ | Travel time of mode $m \in \mathcal{M}$ through a distance, considering the traveling speed of the mode |
| $t_m^{df}(\cdot)$ | Travel time of the feeder to access mode $m \in \mathcal{M}$ through a distance, considering the traveling speed of the feeder |
| $t_{mi}^f$ | Feeder travel time to access mode $m \in \mathcal{M}$ in zone $i \in \mathcal{Z}$ |
| $t_{mi}^r$ | Travel time of mode $m \in \mathcal{M}$ passing through zone $i \in \mathcal{Z}$ |
| $\mu_i$ | Centroid of cluster zone $i \in \mathcal{Z}$ |

*Zonal connection optimization – MILP decision variables*

| | |
|---|---|
| $f_{om}^\alpha$ | Starting flow originating from zone $o \in \mathcal{Z}$ to its node of mode $m \in \mathcal{M}$ |
| $f_{omd}^\beta$ | Ending flow originating from zone $o \in \mathcal{Z}$, which terminates in zone $d \in \mathcal{Z}$ from its node of mode $m \in \mathcal{M}$ |
| $f_{omij}^\lambda$ | Interzonal flow originating from zone $o \in \mathcal{Z}$, which travels from a node of mode $m \in \mathcal{M}$ in zone $i \in \mathcal{Z}$ to another node of mode $m$ in zone $j \in \mathcal{Z}$ |
| $f_{om_1 m_2 i}^\chi$ | Intermodal transfer flow originating from zone $o \in \mathcal{Z}$, which transfers in zone $i \in \mathcal{Z}$ from a node of mode $m_1 \in \mathcal{M}$ to a node of mode $m_2 \in \mathcal{M}$ |
| $x_{mij}^\lambda$ | Integer zonal connection variable that indicates how many direct physical links of transit mode $m \in \mathcal{M}^t$ are used from zone $i \in \mathcal{Z}$ to zone $j \in \mathcal{Z}$ |
| $x_{mij}^\omega$ | Integer infrastructure variable that indicates how many new direct physical links of infrastructure mode $m \in \mathcal{M}^i$ are built from zone $i \in \mathcal{Z}$ to zone $j \in \mathcal{Z}$ |

*Route generation – MILP decision variables*

| | |
|---|---|
| $f_{oijmu}^\nu$ | Interzonal flow, originating from zone $o \in \mathcal{Z}$, which travels on route segment $u \in \mathcal{U}_{mij}$ of transit mode $m \in \mathcal{M}^t$ from zone $i \in \mathcal{Z}$ to zone $j \in \mathcal{Z}$ |
| $f_{omiu}^\psi$ | Starting flow entering route segment $u \in \mathcal{U}_{mij}, j \in \mathcal{Z}\|(i,j) \in \mathcal{L}_m^\lambda$ of transit mode $m \in \mathcal{M}^t$, originating from zone $o \in \mathcal{Z}$ |
| $f_{omju}^\eta$ | Ending flow leaving route segment $u \in \mathcal{U}_{mij}, i \in \mathcal{Z}\|(i,j) \in \mathcal{L}_m^\lambda$ of transit mode $m \in \mathcal{M}^t$, originating from zone $o \in \mathcal{Z}$ |
| $f_{omju_1u_2}^\delta$ | Direct route flow between route segment $u_1 \in \mathcal{U}_{mij}$ which is directly connected to $u_2 \in \mathcal{U}_{mjk}$ in zone $j$ of transit mode $m \in \mathcal{M}^t$, where $i,k \in \mathcal{Z}\|(i,j),(j,k) \in \mathcal{L}_m^\lambda$ |
| $f_{omju_1u_2}^\tau$ | Transfer flow between route segment $u_1 \in \mathcal{U}_{mij}$ which is *not* directly connected to $u_2 \in \mathcal{U}_{mjk}$ in zone $j$ of transit mode $m \in \mathcal{M}^t$, where $i,k \in \mathcal{Z}\|(i,j),(j,k) \in \mathcal{L}_m^\lambda$ |
| $x_{mju_1u_2}^\delta$ | Binary route segment connection variable that $x_{mju_1u_2}^\delta = 1$ if the route segment $u_1 \in \mathcal{U}_{mij}$ is directly connected to $u_2 \in \mathcal{U}_{mjk}$ in zone $j \in \mathcal{Z}$ for transit mode $m \in \mathcal{M}^t$, where $i,k \in \mathcal{Z}\|(i,j),(j,k) \in \mathcal{L}_m^\lambda$; $x_{mju_1u_2}^\delta = 0$ otherwise |

## 3.2 Zonal modeling

Figure 3 illustrates the network elements. Trip origins and destinations are in zones $i \in \mathcal{Z}$ and $j \in \mathcal{Z}$ which are served by interzonal transit or SAMS. A link $(i,j) \in \mathcal{L}_m$ is the fundamental physical unit of a network, representing a connection from zone $i$ to $j$ with a mode $m \in \mathcal{M}$. A link can only be used if it is active, i.e., $(i,j) \in \mathcal{L}_m^\lambda$. For an infrastructure mode $m \in \mathcal{M}^i$, a link is active if it pre-exists or is constructed; for other modes, a link can be used anytime. A route $r \in \mathcal{R}_m$ travels along enabled links $(i,j), (j,k), \ldots \in \mathcal{L}_m^\lambda$ with a segment from zone $i \in \mathcal{Z}$ to $j \in \mathcal{Z}$ denoted as $u \in \mathcal{U}_{mij}$ and a segment of route $r$ as $u \in \mathcal{U}_r$.

The model considers complete trips, which encompass starting (walking or feeder and waiting), interzonal travel (transit modes/SAMS), transfer (intermodal in zonal connection optimization and intramodal in route generation), and ending (walking or feeder). To start a trip, travelers access an interzonal mode by walking[1] or feeders, depending on their proximity to the station of the interzonal mode. Then, they will get on the interzonal mode and travel to another zone, possibly through other zones with the same or other modes. Finally, they will get off the interzonal mode and access their final destinations by walking/active mobility or feeders, depending on the distance. This model only considers *interzonal* trips that originate and terminate in different zones, but not intrazonal trips.

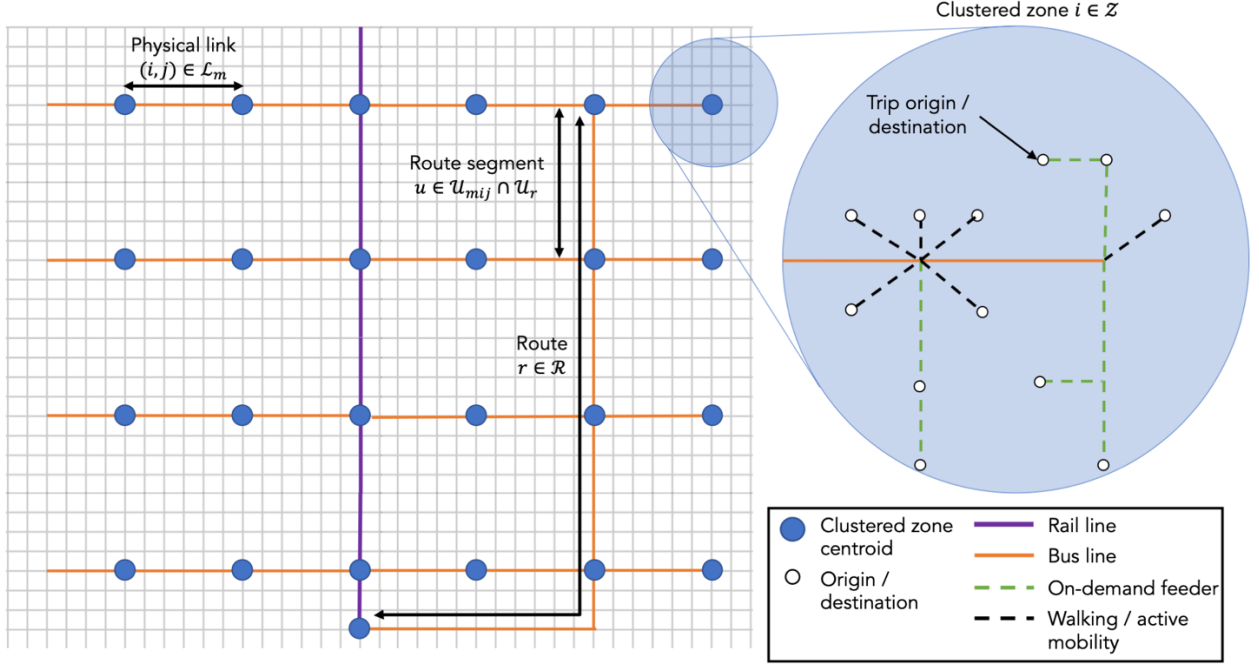

Figure 3. Illustration of multimodal network elements.

### 3.2.1  *k*-means clustering

The goal of the *k*-means geospatial clustering is to cluster the demand into representative zones. Possible alternative approaches include using existing transit regions in a city, consideration of existing transit connections, and geographical boundaries (e.g., hills and islands). To minimize the distances for travelers in MAZ $a \in \mathcal{A}_i$ in each zone $i \in \mathcal{Z}$ to the corresponding zone centroids $\mu_i$, this study employs *k*-means clustering algorithm with a weighting of the squared number of trips of each demand point $(N_a^t)^2$ in Eq. (1) and $|\mathcal{Z}|$ representing the number of zones.

$$\min_{\mu_i} \sum_{i=1}^{|\mathcal{Z}|} \sum_{a \in \mathcal{A}_i} (N_a^t)^2 \|P_a - \mu_i\|_2^2 \tag{1}$$

---

[1] Walking is the assumed form of active mobility here, but other options, such as cycling and scooters, can be readily included by adjusting the parameters $T_m^a$ and $V^a$.

The zone boundary is defined with the Voronoi diagram which partitions a plane into zones closest to the cluster centroids. This helps to translate existing transit networks into the zonal representation. To limit the possible combinations of zone connection from all $(\mathcal{Z} \times \mathcal{Z})$ to only promising transit/SAMS services, the model considers a limited number of potential links. This leads to a much smaller solution space for determining which link $(i,j) \in \mathcal{L}_m$ to enable in the connection optimization step.

Links are added to $\mathcal{L}_m$ based on the following criteria:
1. Proximity: Each zone $i \in \mathcal{Z}$ is connected with its interfacing zones in the Voronoi diagram and its nearest $N^a$ zones $j \in \mathcal{Z}$ by Euclidean distance $d^r_{mij}$. This represents local transit services across zones.

2. Demand: Each zone $i \in \mathcal{Z}$ is connected with another $N^d$ zones $j \in \mathcal{Z}$ with the highest bi-directional O-D $(E_{ij} + E_{ji})$. This includes express transit services skipping zones.

3. Direct transit connection: Each zone $i \in \mathcal{Z}$ is connected with another zone $j \in \mathcal{Z}$ if there is an existing direct transit (e.g., rail) service between the two zones (if existing transit lines are considered in the model).

### 3.2.2 Cost computation

Figure 4 demonstrates how trip generation, termination, transfer, and interzonal travel are modeled with multimodal network representation. In zone $i \in \mathcal{Z}$, a central node and modal nodes of mode $m \in \mathcal{M}$ are connected for trip generation and termination with costs $c^\alpha_{mi}$ and $c^\beta_{mi}$, respectively. Every two modal nodes $m_1, m_2 \in \mathcal{M}$ in a zone are connected for intermodal transfer with a cost $c^\chi_{m_1 m_2 i}$. If the zonal connection between zones $i \in \mathcal{Z}$ and $j \in \mathcal{Z}$ is enabled for mode $m \in \mathcal{M}$, the corresponding modal nodes in the two zones are connected for interzonal trips with a cost of $c^\lambda_{mij}$.

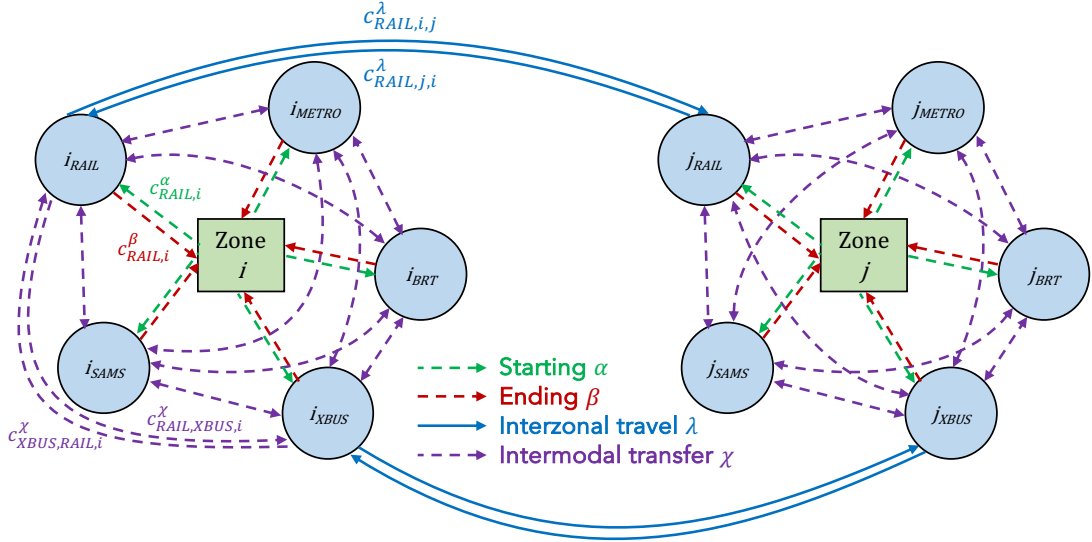

Figure 4. Cost notation in the multimodal network.
*(not all connections shown; flow notation similar)*

This subsection discusses how to compute the four types of arc costs in the multimodal network for each zone and mode from average travel times and distances, based on the geometry and mode characteristics in zones. Geospatial parameters include length and distance; modal parameters include speed, station spacing, and dwell time.[2] The continuous approximation assumes a uniform distribution of demand in a zone, except for the length $L_i$ of each zone defined as the average distance for travelers in zone $i \in \mathcal{Z}$ to leave the zone.

First, we compute the average distance and time in each zone. The traveling time in the Euclidean distance between the boundaries of two zones $t^c_{mij}$ is approximated as a time function[3] $t^d_m(\cdot)$ of the side-to-side distance $D^c_{ij}$ in Eq. (2).

$$t^c_{mij} = t^d_m(D^c_{ij}), \forall (i,j) \in \mathcal{L}_m, \forall m \in \mathcal{M} \tag{2}$$

Similarly, the travel time to pass through a zone from boundary to boundary $t^r_{mi}$, is approximated as a function of the average center-to-side distance $L_i$ with detour $\phi_m$ and dwell time $T^s_m$ with design stop spacing $D^s_m$ in Eq. (3).

$$t^r_{mi} = t^d_m(\phi_m L_i) + \frac{L_i}{D^s_m} T^s_m, \forall i \in \mathcal{Z}, \forall m \in \mathcal{M} \tag{3}$$

For travelers to access interzonal modes, two options are assumed as illustrated in Figure 5: (1) walking to/from transit stations if their origins/destinations are within direct walking coverage (a distance of $V^a T^a_m$); (2) taking door-to-door feeder shuttles[4] to transit station otherwise. Eq. (4) estimates the proportion of riders who use a feeder to access a transit mode $m \in \mathcal{M}^t$ in zone $i \in \mathcal{Z}$, $p^f_{mi}$, by evaluating the area of direct walking coverage (a square of length $2\sqrt{2}V^a T^a_m$) over the catchment area of a station ($L_i D^s_m$). For SAMS, $p^f_{SAMS,i} = 1$ such that all passengers are picked up/dropped off at their origin/destination.

$$p^f_{mi} = \begin{cases} 1 - \frac{8(V^a T^a_m)^2}{L_i D^s_m} \geq 0, m \in \mathcal{M}^t \\ 1, m = SAMS \end{cases}, \forall i \in \mathcal{Z} \tag{4}$$

The average travel distance on feeders $d^f_{mi}$ is then estimated to be a quarter of the zone length, assuming the interzonal route runs through the middle of the zone, with detours in Eq. (5). Subsequently, Eq. (6) estimates travel time $t^f_{mi}$ with the function $t^{df}_m(\cdot)$ of feeder distance and dwell time proportional to the average numbers of boarding/alighting a traveler has to wait for $((R^{df}_m - 1)/2)$.

$$d^f_{mi} = \frac{\phi^f_m L_i}{4}, \forall i \in \mathcal{Z}, \forall m \in \mathcal{M} \tag{5}$$

---

[2] Specifically for SAMS with reference to Section 2.3, its autonomous driving capability is attributed to the lower operating costs $\gamma^o_{SAMS}$, shorter waiting time $T^w_{SAMS}$ and higher utilization rate $\rho_{SAMS}$, while its shared aspect is expected to increase occupancy $R^d_{SAMS}$ and detours $\phi_{SAMS}$ on top of the reduced costs and waiting time.

[3] The time function can incorporate real-world traffic time or microsimulation results incorporating impacts of redesigned transit network and SAMS. For demonstration, this paper adopts constant speeds $V^r_m$ and $V^{rf}_m$ for interzonal services and feeders respectively, and differently in the city and suburbs.

[4] The study assumes SAMS feeder services, but the framework applies to more traditional feeders, e.g., local buses or demand-responsive transit.

$$t_{mi}^f = t_m^{df}(d_{mi}^f) + \frac{(R_m^{df}-1)}{2}T_m^{sf}, \forall i \in \mathcal{Z}, \forall m \in \mathcal{M} \tag{6}$$

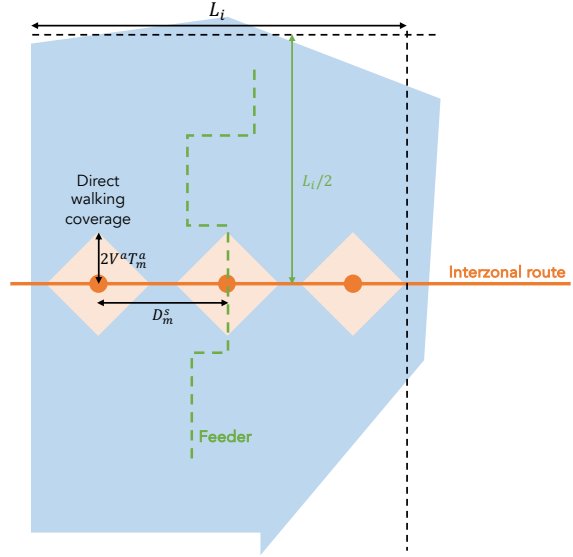

Figure 5. Illustration of walking and feeders for the main route

With the distances and times, we can compute the costs for each zone and mode as multiples of traveling time. Eq. (7) calculates the starting cost $c_{mi}^\alpha$ with walking cost in the first term, feeder costs in the second term (waiting for and traveling on feeders), and waiting costs for the interzonal services. Cost multipliers for walking $\gamma^a$ and waiting $\gamma^w$ are applied to reflect their disutility to travelers relative to in-vehicle travel time. Waiting times $T_m^w$ and $T_m^{wf}$ are estimated as half of the design headway for transit modes, i.e., $T_m^w = H_m^d/2, \forall m \in \mathcal{M}^t$, and pre-set parameters for SAMS[5]. Eq. (8) calculates the ending cost $c_{mi}^\beta$ similarly except without the waiting costs for the interzonal trips.

$$c_{mi}^\alpha = (1-p_{mi}^f)\gamma^a T_m^a + p_{mi}^f(\gamma^w T_m^{wf} + t_{mi}^f) + \gamma^w T_m^w, \forall i \in \mathcal{Z}, \forall m \in \mathcal{M} \tag{7}$$

$$c_{mi}^\beta = (1-p_{mi}^f)\gamma^a T_m^a + p_{mi}^f(\gamma^w T_m^{wf} + t_{mi}^f), \forall i \in \mathcal{Z}, \forall m \in \mathcal{M} \tag{8}$$

The intermodal transfer costs $c_{m_1 m_2 i}^\chi$ from mode $m_1 \in \mathcal{M}$ to $m_2 \in \mathcal{M}$ are calculated with the transfer time in between and waiting costs for mode $m_2$ in Eq. (9).

$$c_{m_1 m_2 i}^\chi = \gamma^a T_{m_1 m_2}^t + \gamma^w T_{m_2}^w, \forall i \in \mathcal{Z}, \forall m_1, m_2 \in \mathcal{M} \tag{9}$$

Eq. (10)-(11) calculate the interzonal distance $d_{mij}^\lambda$ and cost $c_{mij}^\lambda$ traveling from zone $i \in \mathcal{Z}$ to $j \in \mathcal{Z}$ for mode $m \in \mathcal{M}$ by considering the through distance (half of zone length) to travel in each zone with detours and the side-to-side distance between the two zones $D_{ij}^c$.

---

[5] There is no actual transfer between SAMS and the virtual "feeders", as in reality travelers will be served with one single SAMS trip. This is modeled by setting zero waiting time for SAMS feeders, i.e., $T_{SAMS}^{wf} = 0$.

$$d^\lambda_{mij} = \frac{L_i + L_j}{2}\phi_m + D^c_{ij}, \forall (i,j) \in \mathcal{L}_m, m \in \mathcal{M} \tag{10}$$

$$c^\lambda_{mij} = \frac{t^r_{mi}+t^r_{mj}}{2} + t^c_{mij}, \forall (i,j) \in \mathcal{L}_m, \forall m \in \mathcal{M} \tag{11}$$

## 3.3 Zonal connection optimization

The zonal connection optimization model determines the system-optimal infrastructure investment and modes serving flows between zones. Formulated as an MILP problem, the MCNF model considers the characteristics of zone geometry and modes through costs computed in Section 3.2, determines the optimal modal infrastructure investment, and solves for flows to minimize the total generalized costs to serve the demand in the network.

In the multimodal network shown in Figure 4, intermodal transfers[6] are possible through links between modal nodes in each zone. The existing transit modal networks are inserted as interzonal links. Express services are separated from local services by links directly connecting two zones to avoid additional travel costs through intermediate zones.

The commodities in the MCNF model are origins to ensure O-D demand satisfaction. This results in a flow tree for each origin that allows O-D flow back-tracing. Determining the link flows for each O-D is straightforward by solving a single-origin linear program constrained to the model results.

The decision variables of the MILP consist of (1) continuous flow variables $f^\alpha_{om}, f^\beta_{omd}, f^\lambda_{omij}, f^\chi_{om_1m_2i}$, which are labeled by origins and modes, (2) integer zonal connection variables $x^\lambda_{mij}$, and (3) integer infrastructure variables $x^\omega_{mij}$. The last two integer variable groups indicate the existence of physical modal links and set upper bounds for the flow variables. The MILP is solved by determining the optimal combinations of link investments and connections and then the subsequent flows, which allow evaluation of the costs in the objective.

The objective function in Eq. (12) aims to minimize the total door-to-door generalized costs for travelers and operators over all origin zones $o \in \mathcal{Z}$ and mode $m \in \mathcal{M}$, subject to constraints (13)-(29). The first term of the objective is the travelers' costs for starting (access, feeder, and waiting), interzonal travel, intermodal transfer, and ending (access and feeder). Each cost is evaluated by the products of flows and unit costs computed previously.

The second term is the interzonal operating costs by inferring the number of vehicles required from the link flows $f^\lambda_{omij}$ and occupancy $R^d_m$ of each mode. The operating costs are then computed with the time-based cost coefficients $\gamma^o_m$ and the interzonal travel times $c^\lambda_{mij}$, while the emissions cost coefficients $\gamma^e_m$ are distance-based.

Similarly, the third term is the feeder costs for trips starting and ending in zone $o \in \mathcal{Z}$ for operators by inferring the number of feeder vehicles required from the flows of starting $f^\alpha_{om}$ and ending $f^\beta_{imo}$, feeder usage proportion $p^f_{mo}$, and design occupancy $R^{df}_m$.

---

[6] The costs for *intramodal* transfers (e.g., from one bus route to another) are ignored in zonal connection optimization and addressed in route generation in Section 3.4.

$$\min_{\substack{f^\alpha_{om}, f^\beta_{omd}, f^\lambda_{omij}, f^\chi_{om_1m_2i} \\ x^\lambda_{mij}, x^\omega_{mij}}} \sum_{o \in Z} \sum_{m \in \mathcal{M}} \left[ \gamma^r \left( f^\alpha_{om} c^\alpha_{mo} + \sum_{(i,j) \in \mathcal{L}_m} f^\lambda_{omij} c^\lambda_{mij} \right. \right.$$

$$+ \sum_{m_1 \in \mathcal{M}} \sum_{i \in Z} f^\chi_{omm_1 i} c^\chi_{mm_1 i} + \sum_{d \in Z} f^\beta_{omd} c^\beta_{md} \Bigg)$$

$$+ \sum_{(i,j) \in \mathcal{L}_m} \frac{f^\lambda_{omij}}{R^d_m} \left( \gamma^o_m c^\lambda_{mij} + \gamma^e_m d^\lambda_{mij} \right)$$

$$+ \left( f^\alpha_{om} + \sum_{i \in Z} f^\beta_{imo} \right) \frac{p^f_{mo}}{R^{df}_m} \left( \gamma^{of}_m t^f_{mo} + \gamma^{ef}_m d^f_{mo} \right) \Bigg] \quad (12)$$

s.t. (13)-(29)

The first batch of constraints stipulates the bounds of the decision variables. The flow variables for starting $f^\alpha_{om}$, ending $f^\beta_{omd}$, interzonal travel $f^\lambda_{omij}$, and intermodal transfer $f^\chi_{om_1m_2i}$ are set as non-negative in Eq. (13)-(16). Eq. (14) also sets no ending flow at the origin. Eq. (15) avoids flows outside the considered network $\mathcal{L}_m$ and self-loops in the origin zone. Eq. (16) avoids self-loops for intermodal transfers. Eq. (17) stipulates the zonal connection variables $x^\lambda_{mij}$ and infrastructure variables $x^\omega_{mij}$ to be non-negative integers.

$$f^\alpha_{om} \geq 0, \forall o \in Z, m \in \mathcal{M} \quad (13)$$

$$\begin{aligned} f^\beta_{omd} &\geq 0, d \neq o \\ f^\beta_{omd} &= 0, d = o \end{aligned}, \forall o, d \in Z, m \in \mathcal{M} \quad (14)$$

$$\begin{aligned} f^\lambda_{omij} &\geq 0, o \neq j \text{ and } (i,j) \in \mathcal{L}_m \\ f^\lambda_{omij} &= 0, o = j \text{ or } (i,j) \neq \mathcal{L}_m \end{aligned}, \forall o, i, j \in Z, m \in \mathcal{M} \quad (15)$$

$$\begin{aligned} f^\chi_{om_1m_2i} &\geq 0, m_1 \neq m_2 \\ f^\chi_{om_1m_2i} &= 0, m_1 = m_2 \end{aligned}, \forall o, i \in Z, m_1, m_2 \in \mathcal{M} \quad (16)$$

$$\begin{aligned} 0 \leq x^\lambda_{mij} &\in \mathbb{Z}, \forall (i,j) \in \mathcal{L}_m, \forall m \in \mathcal{M}^t \\ 0 \leq x^\omega_{mij} &\in \mathbb{Z}, \forall (i,j) \in \mathcal{L}_m, \forall m \in \mathcal{M}^i \end{aligned} \quad (17)$$

The following four constraints are for flow balance in the multimodal network. Eq. (18) and (19) ensure the flow balance at starting and termination nodes such that the sum of starting and ending flows equal the O-D demand, i.e., serving all demand. Eq. (20) governs the flow balance at origin modal nodes, where there is no intermodal transfer, or inbound or terminating flow. Eq. (21) governs the flow balance at modal nodes other than the origin, such that the inbound flows (from other nodes or modes) are equal to the sum of outbound flows (to other nodes or modes) and ending flows.[7]

$$\sum_{m \in \mathcal{M}} f_{om}^{\alpha} = \sum_{d \in \mathcal{Z}} E_{od}, \forall o \in \mathcal{Z} \tag{18}$$

$$\sum_{m \in \mathcal{M}} f_{omd}^{\beta} = E_{od}, \forall o, d \in \mathcal{Z} \tag{19}$$

$$f_{om}^{\alpha} - \sum_{j|(o,j) \in \mathcal{L}_m} f_{omoj}^{\lambda} = 0, \forall o \in \mathcal{Z}, m \in \mathcal{M} \tag{20}$$

$$\sum_{j|(j,i) \in \mathcal{L}_m} f_{omji}^{\lambda} + \sum_{m_1 \in \mathcal{M}} f_{om_1mi}^{\chi} - \sum_{j|(i,j) \in \mathcal{L}_m} f_{omij}^{\lambda} - \sum_{m_2 \in \mathcal{M}} f_{omm_2i}^{\chi} - f_{omi}^{\beta} = 0, i \neq o, \forall i, o \in \mathcal{Z}, m \in \mathcal{M} \tag{21}$$

Eq. (22) is the infrastructure investment budget constraint. The total infrastructure cost is the sum of zonal infrastructure variables $x_{mij}^{\omega}$ multiplied by the distance $d_{mij}^{r}$ and distance-based unit cost $\gamma_m^i$ across all modes with infrastructure investment $m \in \mathcal{M}^i$. Eq. (23) ensures that sufficient new links are built given the numbers of required links $x_{mij}^{\lambda}$ and existing links $N_{mij}^{\omega}$.

$$\sum_{m \in \mathcal{M}^i} \sum_{(i,j) \in \mathcal{L}_m | i < j} \gamma_m^i d_{mij}^r x_{mij}^{\omega} \leq \Gamma \tag{22}$$

$$x_{mij}^{\omega} \geq x_{mij}^{\lambda} - N_{mij}^{\omega}, \forall (i,j) \in \mathcal{L}_m, \forall m \in \mathcal{M}^i \tag{23}$$

To mitigate concerns of over-capacity on transit lines, Eq. (24) limits the transit link flow $f_{omij}^{\lambda}$ based on the capacity provided by a link[8], which is approximated by the number of links $x_{mij}^{\lambda}$, maximum feasible occupancy $R_m^{max}$, and minimum feasible headway $H_m^{min}$. Eq. (25) ensures bi-directional routes by setting the same number of links for both directions.[9]

$$\sum_{o \in \mathcal{Z}} f_{omij}^{\lambda} \leq \frac{R_m^{max}}{H_m^{min}} x_{mij}^{\lambda}, \forall (i,j) \in \mathcal{L}_m, \forall m \in \mathcal{M}^t \tag{24}$$

---

[7] Eq. (20) avoids superfluous intermodal transfers at the origins by directly splitting starting flows to interzonal flows. Nevertheless, Eq. (21) cannot serve the same purpose at the destinations as the flows are identified by origins but not also by destinations. In the case of ending cost differential, i.e., transferring to another mode and terminating there costs less than terminating directly at a mode, superfluous intermodal transfers are possible. However, cost differentials are rare with realistic parameters.

[8] If transit link capacity is not a concern, combining $x_{mij}^{\lambda}$ and $x_{mij}^{\omega}$ into a group of binary variables to indicate whether zone $i \in \mathcal{Z}$ to $j \in \mathcal{Z}$ is connected with mode $m \in \mathcal{M}^t$ may bring computational benefits. Big-M constraints should however be added to ensure $f_{omij}^{\lambda} = 0$ when $x_{mij}^{\lambda} = 0$, i.e., $f_{omij}^{\lambda} - M x_{mij}^{\lambda} \leq 0, i \neq j, j \neq o, \forall i, j, o \in \mathcal{Z}, m \in \mathcal{M}^t$, where $M$ is a sufficiently large constant. Additionally, it is possible to set a lower bound for the transit link flows based on the minimum service standard with a maximum policy headway. See Appendix A for the discussion and formulation.

[9] The formulation can be used for unidirectional routes by relaxing constraint (25), but the parameters should be adjusted accordingly to capture the operational changes due to uneven service, e.g., deadheading.

$$x^\lambda_{mij} = x^\lambda_{mji}, \forall (i,j) \in \mathcal{L}_m, \forall m \in \mathcal{M}^t \tag{25}$$

The last four constraints model the capacities of interzonal SAMS in terms of road traffic, pick-ups and drop-offs, and the required number of vehicles. These are upper bounds on the SAV flows: vehicular flows in a link, approximated by the passenger flows on SAVs $f^\lambda_{omij}$ divided by average occupancy $R^d_{SAMS}$, $S^\lambda_{SAMS}$ within practical traffic capacity on road networks in Eq. (26); hourly number of trips originating at a zone $F^{\alpha,max}_{SAMS}$ in and terminating at a zone $F^{\beta,max}_{SAMS}$ within the pick-up / drop-off capacities at origins and destinations (for example in a dense city area) in Eq. (27)-(28) respectively; and the total interzonal SAV fleet requirement, approximated by total SAV travel time ($f^\lambda_{o,SAMS,i,j} c^\lambda_{SAMS,i,j}$) divided by the occupancy $R^d_{SAMS}$ and utilization $\rho_{SAMS}$, within the interzonal SAV fleet size $S^t_{SAMS}$ in Eq. (29).

$$\sum_{o \in \mathcal{Z}} \frac{f^\lambda_{omij}}{R^d_{SAMS}} \leq S^\lambda_{SAMS}, \forall (i,j) \in \mathcal{L}_{SAMS} \tag{26}$$

$$f^\alpha_{o,SAMS} \leq F^{\alpha,max}_{SAMS}, \forall o \in \mathcal{Z} \tag{27}$$

$$\sum_{o \in \mathcal{Z}} f^\beta_{o,SAMS,d} \leq F^{\beta,max}_{SAMS}, \forall d \in \mathcal{Z} \tag{28}$$

$$\sum_{o \in \mathcal{Z}} \sum_{(i,j) \in \mathcal{L}_{SAMS}} \frac{f^\lambda_{o,SAMS,i,j} c^\lambda_{SAMS,i,j}}{R^d_{SAMS} \rho_{SAMS}} \leq S^t_{SAMS} \tag{29}$$

### 3.4 Route generation

The route generation algorithm designs routes to minimize the *intramodal* transfers for each transit mode by leveraging the system-optimal results of paths and flows from the last step. Due to a lack of routing information, the zonal connection optimization considers only *intermodal* transfers and assumes a monolithic modal network without intramodal transfers. By expanding the previous MCNF model and constraining it to the results from zonal connection optimization (flows and paths), this model connects links to form routes while minimizing intramodal transfers.

This subsection continues to introduce setting the number of route segments in each link, followed by the expanded network representation. It then details the MILP model and demonstrates the approach with an example.

#### 3.4.1 Number of route segments

The interzonal link flow from zone $i \in \mathcal{Z}$ to $j \in \mathcal{Z}$ for a mode $m \in \mathcal{M}$ is set as the sum of origin-based flows $f^\lambda_{omij}$ over all origins $o \in \mathcal{Z}$ from previous results, i.e., $\sum_{o \in \mathcal{Z}} f^\lambda_{omij}$. This defines the set of active links $\mathcal{L}^\lambda_m$ of mode $m$ by all links with positive flows in Eq. (30).

$$\mathcal{L}^\lambda_m = \{(i,j) \in \mathcal{L}_m | \sum_{o \in \mathcal{Z}} f^\lambda_{omij} > 0\}, \forall m \in \mathcal{M} \tag{30}$$

In the network of *each transit mode* $m \in \mathcal{M}^t$, a route segment $u \in \mathcal{U}_{mij} \cap \mathcal{U}_r$ introduced in Section 3.2 refers to the portion of route $r \in \mathcal{R}_m$ in a link $(i,j) \in \mathcal{L}^\lambda_m$. The maximum number of routes going through the link, with each carrying a design occupancy $R_m^d$ at a design headway $H_m^d$ would be proportional to their total flow $\sum_{o \in Z} f_{omij}^\lambda$.[10] We round this number up to set the number of route segments $|\mathcal{U}_{mij}|$ in Eq. (31).[11]

$$|\mathcal{U}_{mij}| = \left\lceil \frac{H_m^d}{R_m^d} \sum_{o \in Z} f_{omij}^\lambda \right\rceil, \forall (i,j) \in \mathcal{L}_m^\lambda, \forall m \in \mathcal{M}^t \tag{31}$$

### 3.4.2 Route generation network

The previous multimodal network model of each transit mode is now expanded by decomposing each link $(i,j) \in \mathcal{L}_m^\lambda$ into route segments $u \in \mathcal{U}_{mij}$ in Figure 6. Figure 6(a) shows an example network[12] that the link (2,5) consists of two route segments $u_{251}$ and $u_{252}$, given that $|\mathcal{U}_{m,2,5}| = 2$ from Eq. (31). Besides the origin subscript $o$ in the zonal connection optimization, the flow variables in this subsection are also subscripted with the route segment $u$.

With the route segments, the previous model is converted into a route generation network model in Figure 6(b), where route segments are represented by nodes connected to other route segment nodes and zone central nodes. The central nodes are connected with the route segment nodes to denote *starting* (demand origin and intermodal in-transfers, with flows $f_{omiu}^\psi$) and *ending* (demand termination and intermodal out-transfers, with flows $f_{omju}^\eta$). Each route segment $u \in \mathcal{U}_{mij}$ of a link $(i,j) \in \mathcal{L}_m^\lambda$ carries a flow $f_{oijmu}^\nu$ which sums up to $f_{omij}^\lambda$.

The route segment nodes are inter-connected by *direct* and *transfer* links to represent trips along the same and different routes respectively. When two route segments $u_1 \in \mathcal{U}_{mij}$ and $u_2 \in \mathcal{U}_{mjk}$ are assigned to the same route, indicated by a route segment connection variable $x_{mju_1u_2}^\delta = 1$, all the inter-segment flows would be assigned to the direct link ($f_{omju_1u_2}^\delta$); otherwise, the flows would go to the transfer link ($f_{omju_1u_2}^\tau$). For example, in Figure 6, $u_{121}$ and $u_{252}$ are connected in Zone 2 (as shown by the green line), so $x_{m,2,u_{121}u_{252}}^\delta = 1$, $f_{o,m,2,u_{121},u_{252}}^\delta \geq 0$, and $f_{o,m,2,u_{121},u_{252}}^\tau = 0$, resulting in no intramodal transfer. On the contrary, $u_{121}$ and $u_{231}$ are not connected in Zone 2 (as shown by the grey line), so $x_{m,2,u_{121}u_{231}}^\delta = 0$, $f_{o,m,2,u_{121},u_{231}}^\delta = 0$, and $f_{o,m,2,u_{121},u_{231}}^\tau \geq 0$, resulting in intramodal transfer if there is a flow from Zone 1 through Zone 2 to Zone 3.

---

[10] Given a fixed occupancy and vehicle requirement, there is a trade-off between the number of routes passing through two zones and their headways. For example, a route of 5 min headway may be replaced with two routes of 10 min headway. This would apply to all segments of a route. If a lower occupancy or a higher headway is selected in the model, more routes would be allowed to serve the flow between the two zones.

[11] The routes generated by this model are then used to develop a set of transit patterns with heuristics based on transit service design (e.g., interlining, short-turning), which serves as inputs to the next-step joint transit-SAMS service level optimization-simulation problem to optimize the transit line headway and SAV fleet size. The latter optimization disables unnecessary or redundant transit patterns if optimal headway is set above policy limits. Therefore, this route generation algorithm errs on the side of generating more routes than necessary.

[12] The proposed network and subsequent MILP model could handle both uni- and bi-directional routes. The examples shown are with bi-directional routes for clarity.

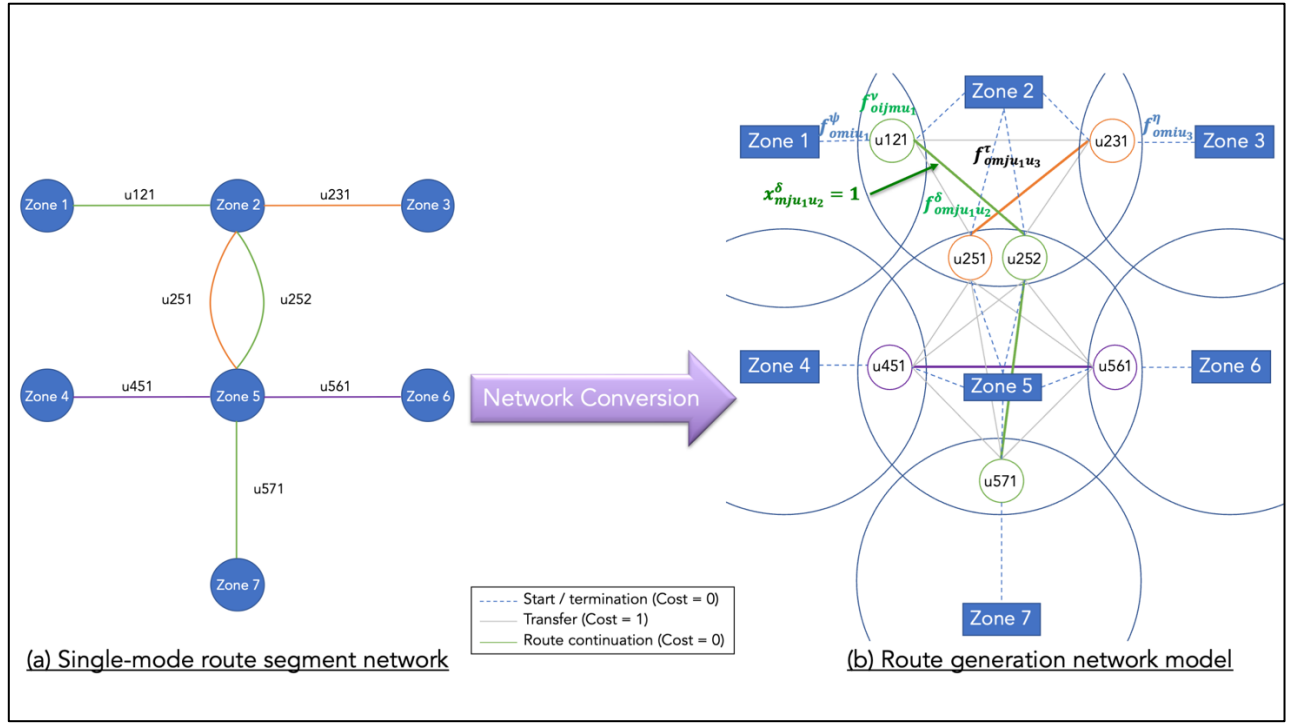

Figure 6. An example of route generation network model

### 3.4.3 MCNF Model

For each transit mode $m \in \mathcal{M}^t$, the route generation MCNF aims to minimize the total number of intramodal transfers $f^\tau_{omju_1u_2}$, summing over all origins, zones, and route segment combinations, subject to constraints (33)-(44). It determines the route segment connections through setting $x^\delta_{mju_1u_2}$ and then solves for the flow variables, the transfer flow of which allows objective evaluation. Different routing would only affect intramodal transfers, but not travel time or operating cost, because paths are set in the zonal connection step.

$$\min_{\substack{x^\delta_{mju_1u_2}, f^\psi_{omiu}, f^\eta_{omju}, \\ f^\nu_{oijmu}, f^\delta_{omju_1u_2}, f^\tau_{omju_1u_2}}} \sum_{(i,j),(j,k)\in\mathcal{L}^\lambda_m} \sum_{\substack{u_1\in\mathcal{U}_{mij} \\ u_2\in\mathcal{U}_{mjk}}} \sum_{o\in\mathcal{Z}} f^\tau_{omju_1u_2}, \forall m \in \mathcal{M}^t \qquad (32)$$

s.t. (33)-(44)

Eq. (33)-(35) set the bounds of decision variables: binary for the route segment connection variable $x^\delta_{mju_1u_2}$ and non-negative for flow variables. Additionally, $x^\delta_{mju_1u_2} = 0$ if $u_1$ and $u_2$ are on the same link in Eq. (33), as represented by $u_1 \in \mathcal{U}_{mij}, u_2 \in \mathcal{U}_{mjk}, i = k$.

$$\begin{cases} x^\delta_{mju_1u_2} \in \{0,1\}, i \neq k \\ x^\delta_{mju_1u_2} = 0, i = k \end{cases}, \forall u_1 \in \mathcal{U}_{mij}, u_2 \in \mathcal{U}_{mjk}, \forall(i,j),(j,k) \in \mathcal{L}^\lambda_m, \forall m \in \mathcal{M}^t \qquad (33)$$

$$f^\psi_{omiu}, f^\eta_{omju}, f^\nu_{oijmu} \geq 0, \forall o \in \mathcal{Z}, u \in \mathcal{U}_{mij}, \forall(i,j) \in \mathcal{L}^\lambda_m, \forall m \in \mathcal{M}^t \qquad (34)$$

$$f^\delta_{omju_1u_2}, f^\tau_{omju_1u_2} \geq 0, \forall o \in \mathcal{Z}, u_1 \in \mathcal{U}_{mij}, u_2 \in \mathcal{U}_{mjk}, \forall(i,j),(j,k) \in \mathcal{L}^\lambda_m, \forall m \in \mathcal{M}^t \qquad (35)$$

The following three constraints govern consistency with the previous zonal connection model results. Eq. (36) ensures consistent start flows with the previous starting demand $f^\alpha_{om}$ at origins or intermodal in-transfers $f^\chi_{om_1mi}$ at non-origins. Eq. (37) ensures consistent end flows with the sum of the previous ending demand $f^\beta_{omj}$ and intermodal out-transfers $f^\chi_{omm_1j}$. Eq. (38) ensures consistent route segment flows with the previous link flows $f^\lambda_{oijm}$.

$$\sum_{j\in Z|(i,j)\in \mathcal{L}^\lambda_m} \sum_{u\in \mathcal{U}_{mij}} f^\psi_{omiu} = \begin{cases} f^\alpha_{om}, & o=i \\ \sum_{m_1\in \mathcal{M}} f^\chi_{om_1mi}, & o\neq i, \end{cases} \forall o,i \in Z, m \in \mathcal{M}^t \quad (36)$$

$$\sum_{i\in Z|(i,j)\in \mathcal{L}^\lambda_m} \sum_{u\in \mathcal{U}_{mij}} f^\eta_{omju} = f^\beta_{omj} + \sum_{m_1\in \mathcal{M}} f^\chi_{omm_1j}, \forall o,j \in Z, m \in \mathcal{M}^t \quad (37)$$

$$\sum_{u\in \mathcal{U}_{mij}} f^\nu_{oijmu} = f^\lambda_{oijm}, \forall o \in Z, (i,j) \in \mathcal{L}^\lambda_m, \forall m \in \mathcal{M}^t \quad (38)$$

Eq. (39) and (40) are flow balance equations inbound and outbound for the route segment $u$ respectively.

$$f^\nu_{ojkmu} = f^\psi_{omju} + \sum_{i\in Z|(i,j)\in \mathcal{L}^\lambda_m} \sum_{u_1\in \mathcal{U}_{mij}} (f^\tau_{omju_1u} + f^\delta_{omju_1u}), \forall o \in Z, u \in \mathcal{U}_{mjk}, \forall (j,k) \in \mathcal{L}^\lambda_m, \forall m \in \mathcal{M}^t \quad (39)$$

$$f^\nu_{oijmu} = f^\eta_{omju} + \sum_{k\in Z|(j,k)\in \mathcal{L}^\lambda_m} \sum_{u_1\in \mathcal{U}_{mjk}} (f^\tau_{omjuu_1} + f^\delta_{omjuu_1}), \forall o \in Z, u \in \mathcal{U}_{mij}, \forall (i,j) \in \mathcal{L}^\lambda_m, \forall m \in \mathcal{M}^t \quad (40)$$

The last four constraints are imposed on the route segment connection variables. Eq. (41) is the Big-M constraint, where $M$ is a sufficiently large constant. The second inequality limits the direct flow $f^\delta_{omju_1u_2}$ to zero if the two route segments are not set as the same route, i.e., $x^\delta_{mju_1u_2}=0$. The first inequality limits $x^\delta_{mju_1u_2}$ to zero if there is no direct flow in both directions at all, i.e., $f^\delta_{omju_1u_2} = f^\delta_{omju_2u_1} = 0, \forall o \in Z$. This avoids excessively long route which serve no through traffic. Eq. (42) ensures bi-directional consistencies[13] of the route segment connection. Eq. (43) and (44) ensure at most one outbound and inbound direct connection respectively to each route segment, i.e., each route segment is connected to at most one segment at either end.

$$-(1-x^\delta_{mju_1u_2}) < \sum_{o\in Z}(f^\delta_{omju_1u_2} + f^\delta_{omju_2u_1}) \leq Mx^\delta_{mju_1u_2}, \forall u_1 \in \mathcal{U}_{mij}, u_2 \in \mathcal{U}_{mjk}, u_1 < u_2, \forall (i,j),(j,k) \in \mathcal{L}^\lambda_m, \forall m \in \mathcal{M}^t \quad (41)$$

$$x^\delta_{mju_1u_2} = x^\delta_{mju_2u_1}, \forall u_1 \in \mathcal{U}_{mij}, u_2 \in \mathcal{U}_{mjk}, \forall (i,j),(j,k) \in \mathcal{L}^\lambda_m, \forall m \in \mathcal{M}^t \quad (42)$$

---

[13] Eq. (41) and (42) cater for bi-directional routes. For uni-directional routes, only one flow $f^\delta_{omju_1u_2}$ should be considered in Eq. (41), and Eq. (42) should be omitted.

$$\sum_{k\in\mathcal{Z}|(j,k)\in\mathcal{L}_m^\lambda}\sum_{u_1\in\mathcal{U}_{mjk}} x^\delta_{mjuu_1} \leq 1, \forall u \in \mathcal{U}_{mij}, \forall (i,j)\in\mathcal{L}_m^\lambda, \forall m \in \mathcal{M}^t \qquad (43)$$

$$\sum_{i\in\mathcal{Z}|(i,j)\in\mathcal{L}_m^\lambda}\sum_{u_1\in\mathcal{U}_{mij}} x^\delta_{mju_1u} \leq 1, \forall u \in \mathcal{U}_{mjk}, \forall (j,k)\in\mathcal{L}_m^\lambda, \forall m \in \mathcal{M}^t \qquad (44)$$

To form routes from the optimization results, we connect route segments $u_1 \in \mathcal{U}_{mij}$ to either end of another route segment $u_2 \in \mathcal{U}_{mjk}$ at zone $j \in \mathcal{Z}$ if $x^\delta_{mju_1u_2} = 1$ iteratively. For example, a route would be $[u_1, u_2, u_3, \dots ]|x^\delta_{u_1u_2} = x^\delta_{u_2u_3} = \cdots = 1, u_1 \in \mathcal{U}_{mi_1i_2}, u_2 \in \mathcal{U}_{mi_2i_3}, u_3 \in \mathcal{U}_{mi_3i_4}, \dots$, which passes through zones $i_1, i_2, i_3, i_4 \dots$.

### 3.4.4 Example illustration

The results of the MILP formulation are demonstrated with the previous example in Section 3.4.2 and benchmarked with a myopic heuristic, which connects route segments to serve the highest local through flows. The number of route segments is determined with Eq. (31). For example, the link (2,5) consists of two route segments $u_{251}$ and $u_{252}$, given that $|\mathcal{U}_{m,2,5}| = 2$.

In Figure 7(a), myopic heuristics try to add links to routes to serve most through traffic, for example, by connecting the route segment between Zones 5 and 6 to Route 1 (saving 50 transfers) instead of Route 3 (saving 40 transfers). However, when multiple routes are on the same link, it may choose the locally optimal route segment to connect without tracing which route would carry more through flow. Specifically, the route segment between Zones 5 and 7 should be connected to Route 1 (saving 40 transfers) instead of 2 (saving 10 transfers) as shown in Figure 7(b). Figure 7(c) compares the performances showing that the proposed algorithm saves 20 transfers. The performance difference would be more significant for longer routes and intersections with more route options.

Figure 7(d) shows the solution represented in the route generation network model. To minimize the total number of transfers, route segment $u_{571}$ is connected to $u_{252}$ by setting $x^\delta_{m,5,u_{252},u_{571}} = 1$ and $x^\delta_{m,5,u_{251},u_{571}} = 0$ and divert the flow to $f^\delta_{1,m,5,u_{252},u_{571}}$ instead of $f^\tau_{1,m,5,u_{252},u_{571}}$, which reduces the objective. The O-D demand from Zone 1 to 7 would flow through $f^\psi_{1,m,1,u_{121}}, f^\nu_{1,1,2,m,u_{121}}, f^\delta_{1,m,2,u_{121},u_{252}}, f^\nu_{1,2,5,m,u_{252}}, f^\delta_{1,m,5,u_{252},u_{571}}, f^\nu_{1,5,7,m,u_{571}}$, and $f^\eta_{1,m,7,u_{571}}$.

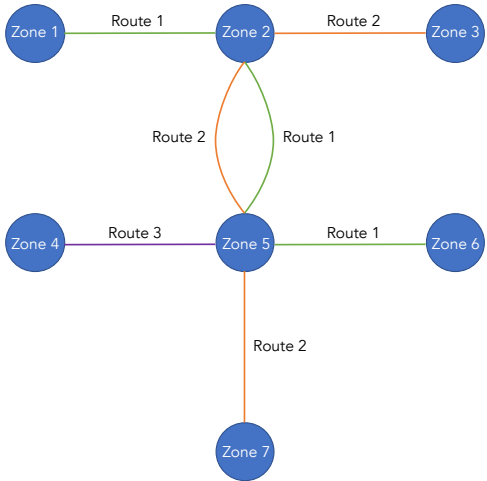
(a) Myopic heuristics

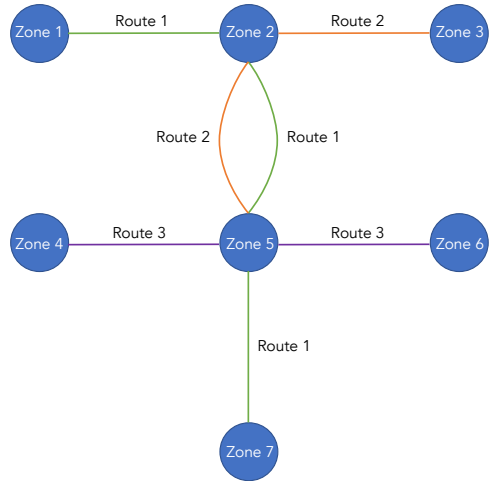
(b) Proposed route generation algorithm

| Origin-destination | Demand | Number of transfers with myopic routing | Number of transfers with the proposed algorithm |
|---|---|---|---|
| 1>5 | 10 | 0 | 0 |
| 1>7 | 40 | 40 | 0 |
| 2>6 | 50 | 0 | 50 |
| 4>6 | 40 | 40 | 0 |
| 3>7 | 10 | 0 | 10 |
| Total | 150 | 80 | 60 |

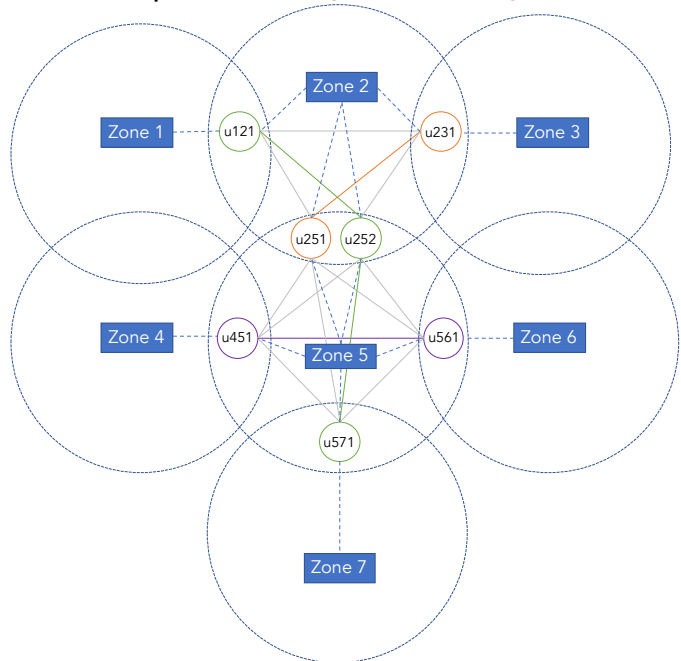

(c) O-D Demand and number of transfers required

(d) Route generation network model representation

Figure 7. Route result example to compare myopic heuristics and the proposed route generation algorithm.

# 4 Experiment Design

The framework efficacy, in terms of optimization performance and solution quality, is demonstrated by designing a large-scale multimodal transit network for the Chicago metropolitan area in the United States. Demand data were adapted from the activity-based demand model CT-RAMP (Chicago Metropolitan Agency for Planning, 2023), where trips originate and end at micro analysis zones (MAZs). Approximately 6,720,000 weekly trips that used transit and taxi[14] are included, further aggregated to 261,432 hourly trips assuming a three-hour daily period to capture peak hour demand pattern.

---

[14] The demand shift from/to private vehicles are not considered in this study.

Subsequently, *k*-means clustering was conducted on the 16,819 MAZs, with their weightings being the squared number of trips originating and ending at each.[15] Figure 8 depicts the results of 50 clustered zones, color-coded to represent the number of trips originating and terminating in each. Zone 22 - Chicago Loop, the Central Business District (CBD), accommodates the highest concentration of trips (nearly 88,300 trips/hour). Neighboring zones within the city boundary (as marked by red lines in Figure 8), such as Zone 31 - Chicago Rogers Park, are urban areas with fewer trips (approximately 20,000 trips/hour). Beyond the city are suburbs like Zone 0 - Joliet, where a declining trip density is observed (around 3,000 trips/hour in the peripheral zone).

Table 2 lists the general parameters and properties of a typical weekday peak service. Unless otherwise noted, the time unit is one hour; for example, all demands and costs are hourly. The model considers seven modes: five transit modes $\mathcal{M}^t$, interzonal SAMS, and walking. Three transit modes, commuter rail (*RAIL*), rapid transit rail (*METRO*), and bus rapid transit (*BRT*), require capital investment or existing infrastructure to run ($\mathcal{M}^i$), while express buses (*XBUS*), local buses (*LBUS*), and interzonal SAMS (*SAMS*) utilize existing road networks. The separation of rail (*RAIL* and *METRO*) and bus services (*XBUS* and *LBUS*) to express and local service classes differentiate the faster yet more difficult-to-access services from the slower services with frequent stops. The operational parameters of transit modes, such as design headway, stop spacing, and occupancies are referenced from design manuals and current operational data (American Public Transportation Association, 2020; Chicago Transit Authority, 2023a). SAMS parameters are assumed with reference to Pinto et al. (2020).

Costing data are obtained from multiple sources reflecting an average U.S. transit system, adjusted for inflation to 2022 US$ using the Producer Price Index in Transportation Industries (U.S. Bureau of Labor Statistics, 2023). The capital costs of transit systems are based on the average of higher-income countries (The Institute of Transportation and Development Policy (ITDP), 2018). The operating costs are the average U.S. revenue-hourly costs from the National Transit Database (Federal Transit Administration, 2021). The greenhouse gas emissions are estimated for U.S. transit systems (Federal Transit Administration, 2017), with a cost of carbon emissions at \$185/tonne $CO_2$ (Rennert et al., 2022). For the user's costs, the value of time is evaluated as 50% of hourly median household income in 2019 in the U.S. (US Department of Transportation, 2016). The coefficients of walking and waiting are set with reference to Wardman (2004).

---

[15] To concentrate clusters to reflect major demand and transit usage, only the urban and suburban areas served by interzonal transit lines are considered in the clustering process. Afterward, the O-D demand from the remaining peripheral areas is added to the nearest clusters to reflect their trips via nearby transit stops.

Figure 8. Clustered zones in the model. *(The Chicago city boundary is denoted by the red line.)*
24

Table 2. Input parameters of the models.

| Sets | |
|---|---|
| $\mathcal{A}$ | $|\mathcal{A}| = 16819$ |
| $\mathcal{L}_m$ | $|\mathcal{L}_m| = 640$ |
| $\mathcal{M}$ | {RAIL, METRO, BRT, XBUS, LBUS, SAMS, WALK} |
| $\mathcal{M}^i$ | {RAIL, METRO, BRT} |
| $\mathcal{M}^t$ | {RAIL, METRO, BRT, XBUS, LBUS} |
| $\mathcal{Z}$ | $|\mathcal{Z}| = 50$ |

| General – parameters | |
|---|---|
| $F_{SAMS}^{\alpha,max}$ | 5000 trips/h |
| $F_{SAMS}^{\beta,max}$ | 5000 trips/h |
| $N^a$ | 10 |
| $N^d$ | 5 |
| $S_{SAMS}^t$ | 20000 veh |
| $S_{SAMS}^\lambda$ | 2000 veh/h |
| $T_{SAMS}^w$ | 5 min |
| $T_{SAMS}^{wf}$ | 0 min |
| $V^a$ | 4 km/h |
| $\gamma^a$ | 2 |
| $\gamma^r$ | $16.5/h |
| $\gamma^w$ | 1.5 |
| $\rho_{SAMS}$ | 0.7 |
| $\Gamma$ | $0 (Scenario 1); $14.88B (Scenario 2); $371.94B (Scenario 3) |

*Modal parameters*

| $m$ | | RAIL | METRO | BRT | XBUS | LBUS | SAMS | WALK |
|---|---|---|---|---|---|---|---|---|
| $D_m^s$ | (km) | 2 | 1.4 | 1 | 1 | 0.5 | / | / |
| $H_m^d$ | (min) | 15 | 5 | 5 | 10 | 5 | / | / |
| $R_m^d$ | (pax/veh) | 600 | 300 | 60 | 40 | 30 | 2 | / |
| $R_m^{df}$ | (pax/veh) | 5 | 5 | 2 | 2 | 2 | 2 | / |
| $T_m^a$ | (min) | 7.5 | 7.5 | 5 | 5 | 5 | / | / |
| $T_m^s$ | (min) | 1 | 1 | 0.75 | 0.75 | 0.5 | 2 | / |
| $T_m^{sf}$ | (min) | 0.5 | 0.5 | 2 | 2 | 2 | 2 | / |
| $T_{m_1 m_2}^t$ | (min) | 5 | 5 | 5 [16] | 5 [16] | 5 | 5 | 5 |
| $V_m^r$(city) | (km/h) | 50 | 50 | 40 | 20 | 20 | 25 | 4 |
| $V_m^r$(suburb) | (km/h) | 80 | 70 | 60 | 50 | 40 | 60 | 4 |
| $V_m^{rf}$(city) | (km/h) | 20 | 20 | 20 | 20 | 20 | 20 | 4 |
| $V_m^{rf}$(suburb) | (km/h) | 40 | 40 | 40 | 40 | 40 | 40 | 4 |
| $\gamma_m^e$ | ($/veh-km) | 3.33 | 3.33 | 1.184 | 0.592 | 0.592 | 0.074 | 0 |
| $\gamma_m^{ef}$ | ($/veh-km) | 0.074 | 0.074 | 0.074 | 0.074 | 0.074 | 0.074 | 0 |
| $\gamma_m^i$ | (M$/km) | 575.37 | 575.37 | 13.34 | / | / | / | / |

---

[16] 0-min transfer costs between *BRT* and *XBUS* to model interlining, i.e., the same *XBUS* route runs on *BRT* guideway.

| | | | | | | | | |
|---|---|---|---|---|---|---|---|---|
| $\gamma_m^o$ | ($/veh-h) | 4764.87 | 2032.82 | 477.39 | 338.51 | 42.9 | 26.4 | 0 |
| $\gamma_m^{of}$ | ($/veh-h) | 26.4 | 26.4 | 26.4 | 26.4 | 26.4 | 26.4 | 0 |
| $\phi_m$ | | 1 | 1.1 | 1.2 | 1.2 | 1.3 | 1.3 | 1 |
| $\phi_m^f$ | | 1.5 | 1.5 | 1.5 | 1.5 | 1.5 | 1.5 | 1 |

*Note: "/" represents not applicable.*

The study examines three scenarios with or without existing rail networks, and with different levels of infrastructure investment budgets. Scenario 1 (*Existing rail networks with no infrastructure investment*) and Scenario 2 (*Existing rail networks with nominal infrastructure investment*) incorporate the existing *RAIL* (756km) and *METRO* (96km) networks and optimize *XBUS* and *LBUS* networks. Additionally, Scenario 2 is allocated a $14.88B infrastructure budget, equivalent to the current level of investment in the transit system, to build new links for *RAIL*, *METRO*, and *BRT*. Scenario 3 (*Greenfield design with no existing transit networks*) optimizes networks of all modes without any existing transit infrastructure but is allocated a $371.94B infrastructure budget, equivalent to building the existing rail infrastructure.[17]

In Scenarios 1 and 2, the existing Chicago transit networks - commuter rail, Metra (City of Chicago, 2012), and rapid transit rail, Chicago "L" (Chicago Transit Authority, 2022) - are integrated into the model with Geographic Information System (GIS) analysis. Figure 9 exhibits their GIS networks ((a) and (c)) along with the interzonal connections captured in the model ((b) and (d)).

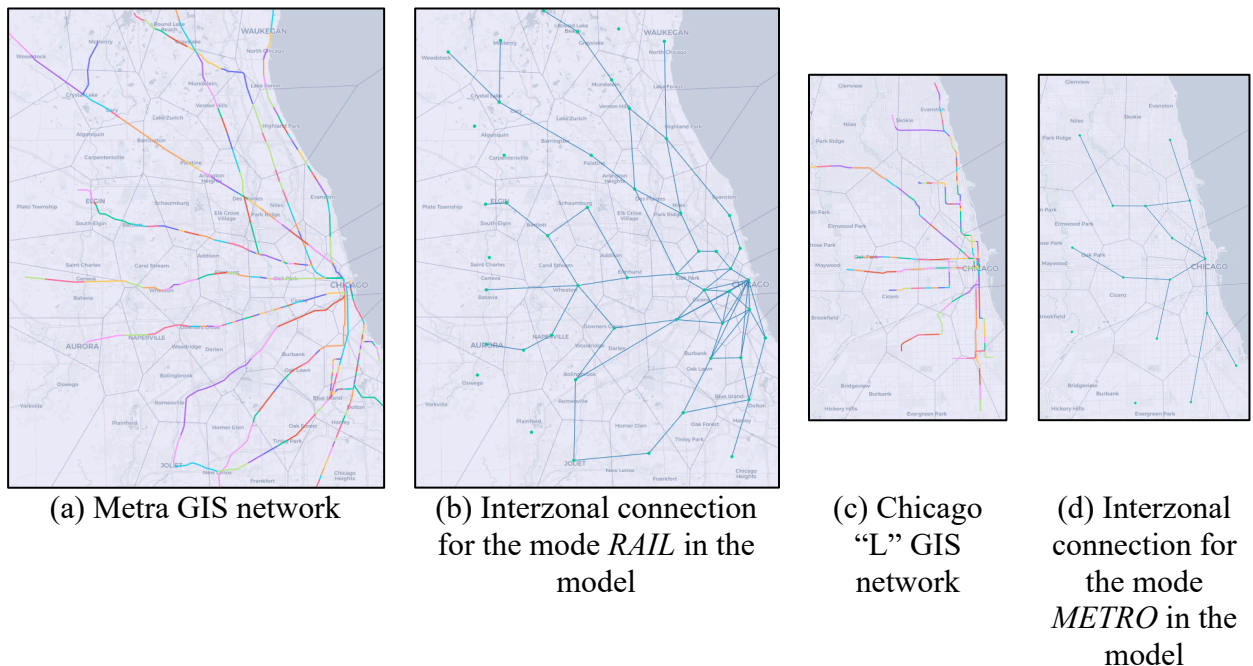

(a) Metra GIS network  (b) Interzonal connection for the mode *RAIL* in the model  (c) Chicago "L" GIS network  (d) Interzonal connection for the mode *METRO* in the model

Figure 9. Existing Chicago train networks.
*(GIS models and interzonal connection in the model of 50 zones)*

---

[17] For Scenario 2, the $14.88B infrastructure investment is equivalent to an annual budget of $1B for 20 years over 3% discount rate, similar to the current level of capital investment in the transit system in Chicago metropolitan area, or 26km of new *RAIL* or *METRO* tracks or 1,116km of BRT guideway in the model. For Scenario 3, the $371.94B budget is also equivalent to 647km of new *RAIL* or *METRO* tracks or 27900km of BRT guideway in the model.

# 5 Computational Results

## 5.1 Zonal connection optimization

The optimization was performed on a computing cluster with the Red Hat Enterprise Linux 7.9 operating system, Intel Xeon Gold 6338 2GHz, 32 cores, and 64 GB RAM. Gurobi 9.5.1 on Python 3.9.12 was used to solve the MILPs.

The MILPs in the three scenarios were solved promptly. In Scenario 1, the MILP is reduced to a linear program because the zero infrastructure budget eliminates discrete options for link addition, so it was solved within seconds. Scenarios 2 and 3 consider diverse combinations of link addition (640 candidate links between 50 zones for each of the 7 modes). Table 3 outlines the results of these scenarios. It is reminded that the results only account for the *interzonal* connections but not the intrazonal trips.

Table 3. Results of zonal connection optimization in the three scenarios.

| Scenario | | 1 (Existing rail networks with no infrastructure investment) | 2 (Existing rail networks with nominal infrastructure investment) | 3 (Greenfield design with no existing rail networks) |
|---|---|---|---|---|
| ***Model characteristics*** | | | | |
| Number of variables: | | | | |
| Continuous: | | 1,015,350 | 1,015,350 | 1,015,350 |
| Integer: | | 20,000 | 20,000 | 20,000 |
| Number of constraints: | | 50,777 | 50,777 | 50,777 |
| ***Computation results*** | | | | |
| Computation time | (h) | 0.001 | 23.00 [18] | 46.00 [18] |
| Objective | ($) | 5,574,873 | 5,486,490 | 5,372,977 |
| Lower bound | ($) | 5,574,873 | 5,404,659 | 5,295,021 |
| Solution gap | | 0.00% | 1.49% | 1.45% |
| ***Scenario results*** | | | | |
| Average generalized cost | ($/trip) | 21.33 | 20.99 | 20.56 |
| Average journey time | (min/trip) | 50.14 | 49.64 | 48.77 |
| Average start time | (min/trip) | 15.34 | 15.23 | 13.12 |
| Average interzonal travel time | (min/trip) | 24.63 | 23.90 | 24.98 |
| Average end time | (min/trip) | 10.06 | 10.37 | 10.54 |
| Average number of intermodal transfers | (/trip) | 0.0094 | 0.0188 | 0.0169 |
| Average operating cost | ($/trip) | 4.81 | 4.58 | 4.46 |

---

[18] 23 and 46 hours are assigned for computation for Scenarios 2 and 3 respectively. To reduce computation time, the transit capacity constraint (24) is omitted for Scenario 3. The results are checked to abide by the original capacity constraint.

| Scenario | | 1 (Existing rail networks with no infrastructure investment) | 2 (Existing rail networks with nominal infrastructure investment) | 3 (Greenfield design with no existing rail networks) |
|---|---|---|---|---|
| Average emissions cost | ($/trip) | 0.32 | 0.31 | 0.37 |
| Modal split by trip count: | | | | |
| *RAIL* | | 46.41% | 42.47% | 1.00% |
| *METRO* | | 35.29% | 33.70% | 70.46% |
| *BRT* | | 0.00% | 14.59% | 27.61% |
| *XBUS* | | 4.47% | 2.55% | 0.10% |
| *LBUS* | | 0.00% | 0.00% | 0.00% |
| *SAMS* | | 13.83% | 6.68% | 0.84% |
| Modal split by distance traveled: | | | | |
| *RAIL* | | 56.95% | 54.19% | 0.77% |
| *METRO* | | 19.52% | 19.25% | 56.89% |
| *BRT* | | 0.00% | 13.68% | 40.65% |
| *XBUS* | | 8.78% | 4.78% | 0.24% |
| *LBUS* | | 0.00% | 0.00% | 0.00% |
| *SAMS* | | 14.75% | 8.09% | 1.45% |
| Average trip distance: | | | | |
| *RAIL* | (km/trip) | 21.05 | 21.26 | / |
| *METRO* | (km/trip) | 9.49 | 9.52 | 13.53 |
| *BRT* | (km/trip) | / | 15.61 | 24.67 |
| *XBUS* | (km/trip) | 33.68 | 31.24 | / |
| *SAMS* | (km/trip) | 18.30 | 20.16 | 29.07 |
| Average speed (including dwell time): | | | | |
| *RAIL* | (km/h) | 47.55 | 47.94 | / |
| *METRO* | (km/h) | 35.67 | 36.17 | 40.47 |
| *BRT* | (km/h) | / | 38.14 | 41.32 |
| *XBUS* | (km/h) | 37.26 | 38.02 | / |
| *SAMS* | (km/h) | 37.80 | 40.43 | 44.15 |
| Average operating cost: | | | | |
| *RAIL* | (¢/pax-km) | 16.70 | 16.57 | / |
| *METRO* | (¢/pax-km) | 18.99 | 18.73 | 16.74 |
| *BRT* | (¢/pax-km) | / | 20.86 | 19.26 |
| *XBUS* | (¢/pax-km) | 22.71 | 22.26 | / |
| *SAMS* | (¢/pax-km) | 34.96 | 32.68 | 29.93 |

*Note: Average results are only shown for modes with modal split more than 1%.*

From Scenario 1 to 3, the model is given more freedom to design the transit network. Accordingly, the generalized costs decrease, while both the average journey time and operating cost are reduced. Compared to Scenario 1, there are 100% and 80% more intermodal transfers respectively in Scenarios 2 and 3, as the model leverages new infrastructure (e.g., BRT) to reduce journey time and operating costs, even after considering transfer penalties. This represents more multimodal cooperation.

The emissions costs are insignificant in contrast to operating costs (less than 10%), despite the relative high unit cost adopted. However, this model does not factor in the emission savings associated with the potential reduction in private vehicle miles traveled.

The following subsections delve into the specific spatial network designs.

### 5.1.1 Scenario 1 - Existing rail infrastructure with no infrastructure investment

This scenario preserves the existing *RAIL* and *METRO* networks and reconfigures the bus networks, further determining the multimodal path for each O-D pair. As Figure 10 indicates, trips within the city are served mainly by rapid transit rail (*METRO*), city-suburb trips by commuter rail (*RAIL*), and suburb-suburb trips by express buses (*XBUS*) and interzonal SAMS. Commuter rail and express buses are mainly used for longer trips, while rapid transit rail is used for shorter trips, which explains the discrepancies in modal splits by trip counts and by distance traveled. Interzonal SAMS account for 15% passenger-km traveled even with the system-optimal objectives because of its benefits for the travelers.

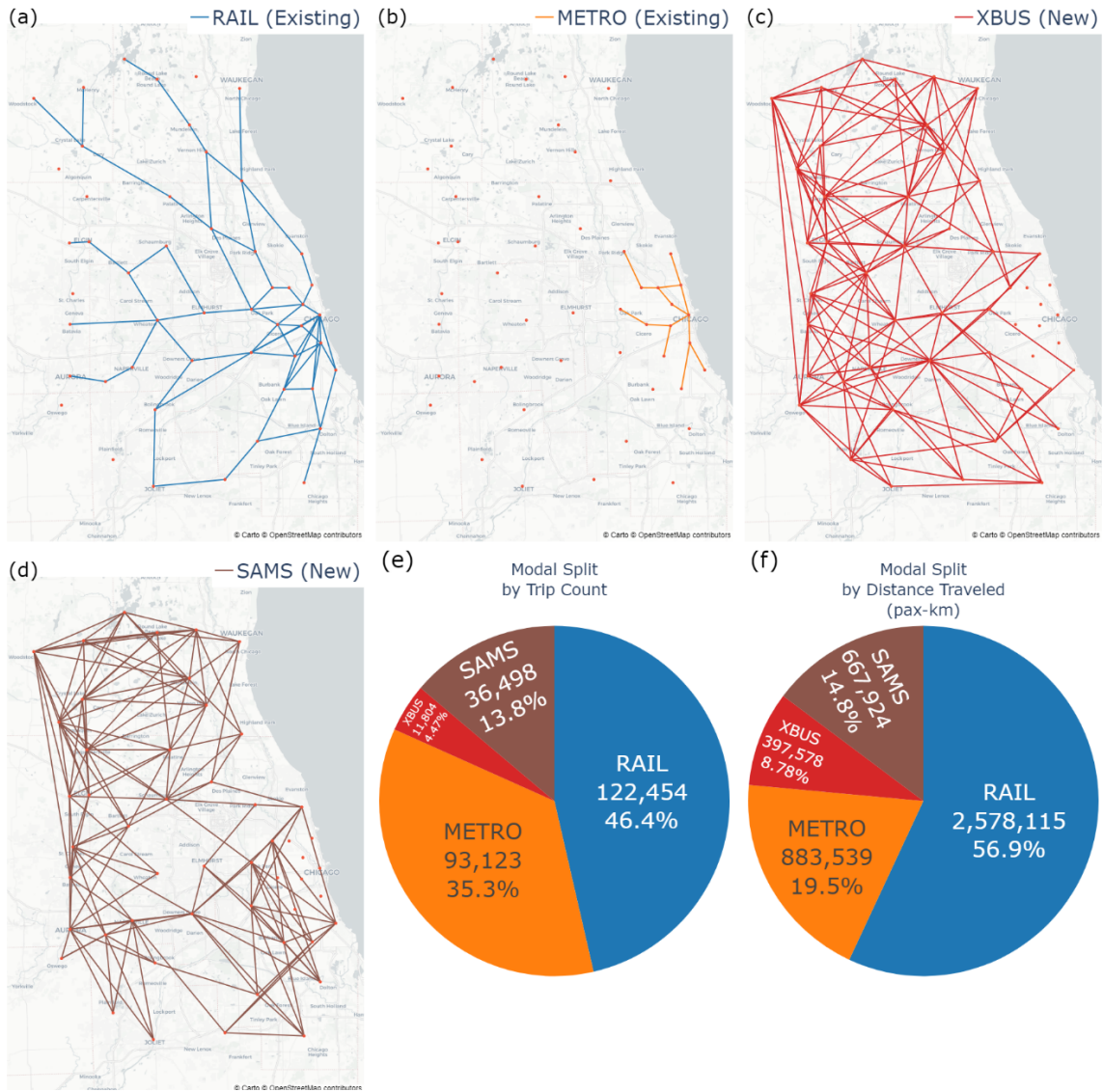

Figure 10. Results of Scenario 1 - Existing rail infrastructure with no new capital investment: (a-d) modal networks, (e) modal split by trip count, (f) modal split by distance traveled.

To highlight the service differences in CBD, urban, and suburban areas, Figure 11 presents the multimodal paths (links colored with the most used mode) and modal splits of trips originating from Chicago Loop, Rogers Park, and Joliet as examples. From the Loop, a CBD, most trips to other parts of the city are served by rapid transit, and suburbs by commuter rail, utilizing the existing networks radiating from the city. There are also some transfers to express buses for reaching more rural suburbs. From Rogers Park, at the city's periphery, more travelers are served with interzonal SAMS and express buses compared to the Loop. This is partly due to the absence of direct rail connections to destinations outside the north-south rail corridor. From Joliet, a suburb, we observe a higher use of interzonal SAMS for shorter trips and express buses for longer trips, differentiating the flexibility and accessibility of the former and the lower costs of the latter.

Overall, more pronounced use of interzonal SAMS is observed in transit gaps, which are some urban-suburban and suburb-suburb pairs without rail connections and the demand may be too sparse to justify regular bus routes. Appendix B features further discussion on modal splits in trips of different distances.

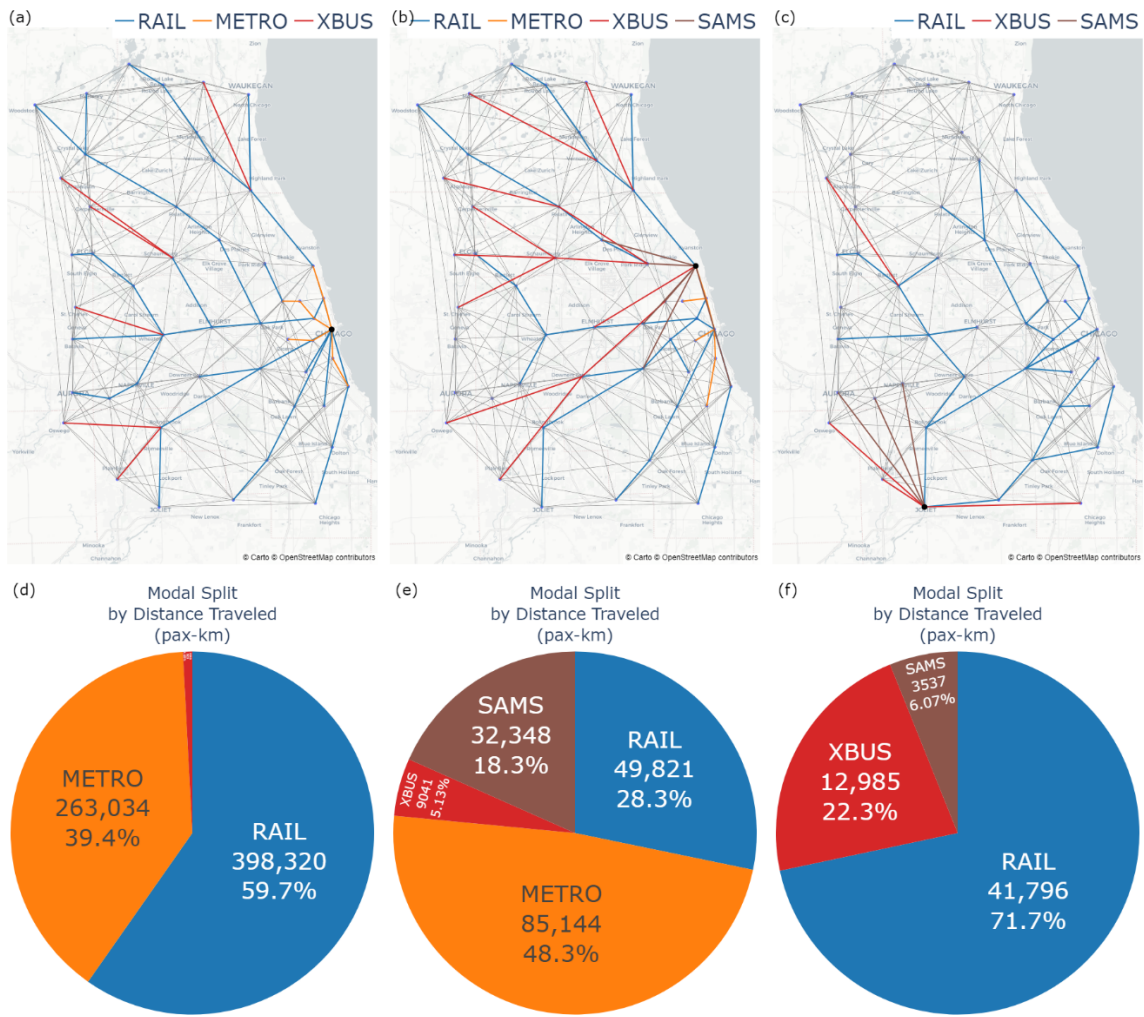

Figure 11. Origin results of Scenario 1 - Existing rail infrastructure with no new capital investment (multimodal paths, modal splits by distance traveled) for trips from: (a,d) Chicago – Loop, (b,e) Chicago – Rogers Park, (c,f) Joliet.

### 5.1.2 Scenario 2 - Existing rail infrastructure with nominal infrastructure investment

Scenario 2 determines *RAIL*, *METRO*, and *BRT* infrastructure investment on top of the existing networks of the first two, and then decides the flow path across all modes. Results are shown in Figure 12. The model introduces a new rapid transit link to provide express services and replace the original local services via another zone. This maintains its market share by distance traveled amid competition from the new BRT system. The new BRT network serves urban-to-suburb and some suburban trips, filling the gaps left by commuter rail and rapid transit and offering more direct trips. It also shrinks the market shares of interzonal SAMS (from 14% in Scenario 1 to 7% in Scenario 2) and express bus (from 5% to 3%) with similarly fast and cheaper services. The redesign with nominal capital investment reduces the average journey time by 0.5min/trip (1%) and operating cost by $0.23/trip (5%).

Figure 13 presents the multimodal paths and mode splits from a CBD, the city periphery, and a suburb. For trips from the Loop, rapid transit gains market share from commuter rail thanks to the newly added link, which saves travel time and operating cost compared to Scenario 1. From Rogers Park, BRT serves more than half interzonal SAMS trips in Scenario 1 and takes away 10% modal split from rapid transit with more direct services. From Joliet, more than 75% express bus trips in Scenario 1 are now upgraded to BRT services.

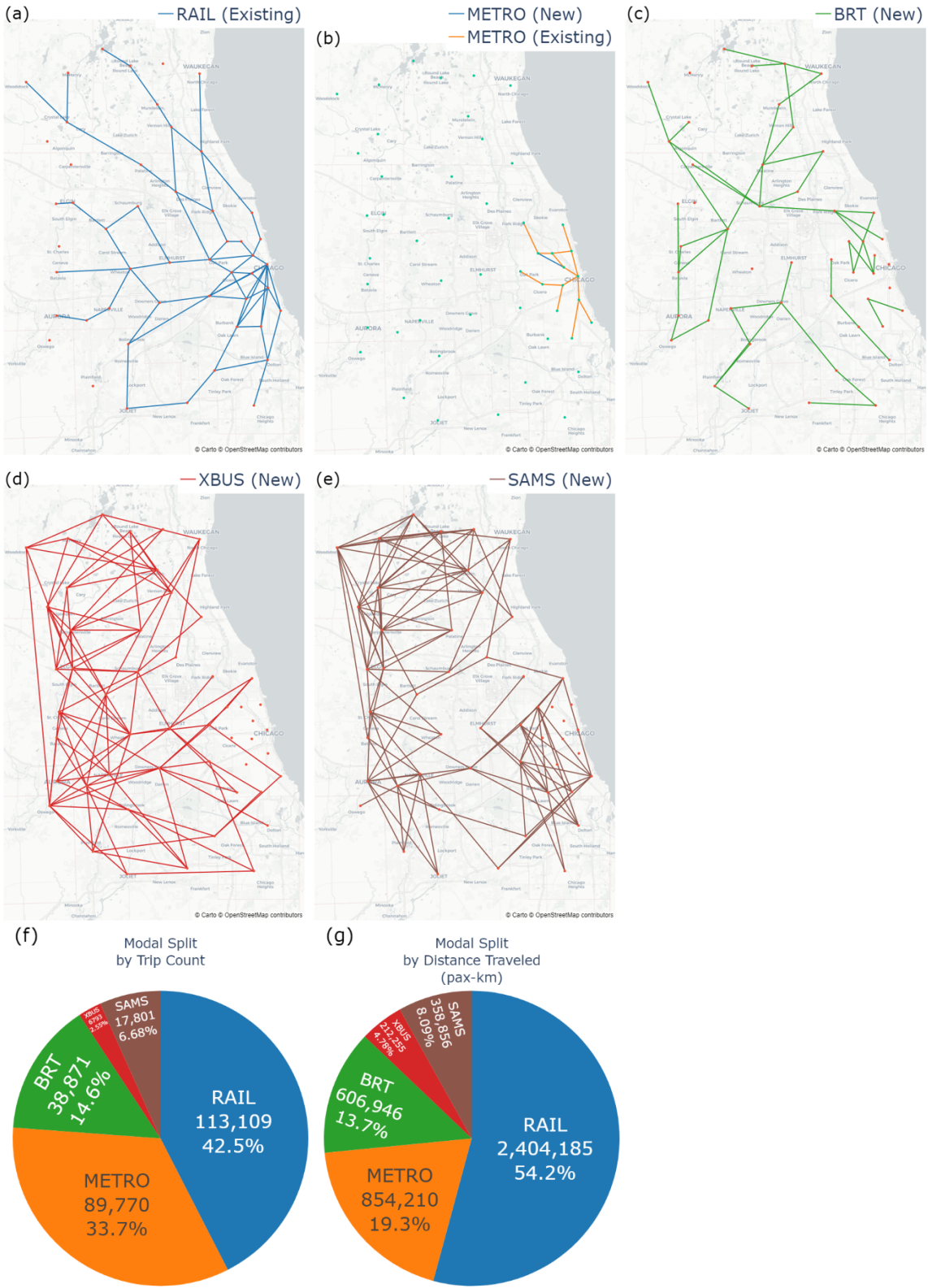

Figure 12. Results of Scenario 2 - Existing rail infrastructure plus nominal capital investment: (a-e) modal networks, (f) modal split by trip count, (g) modal split by distance traveled.

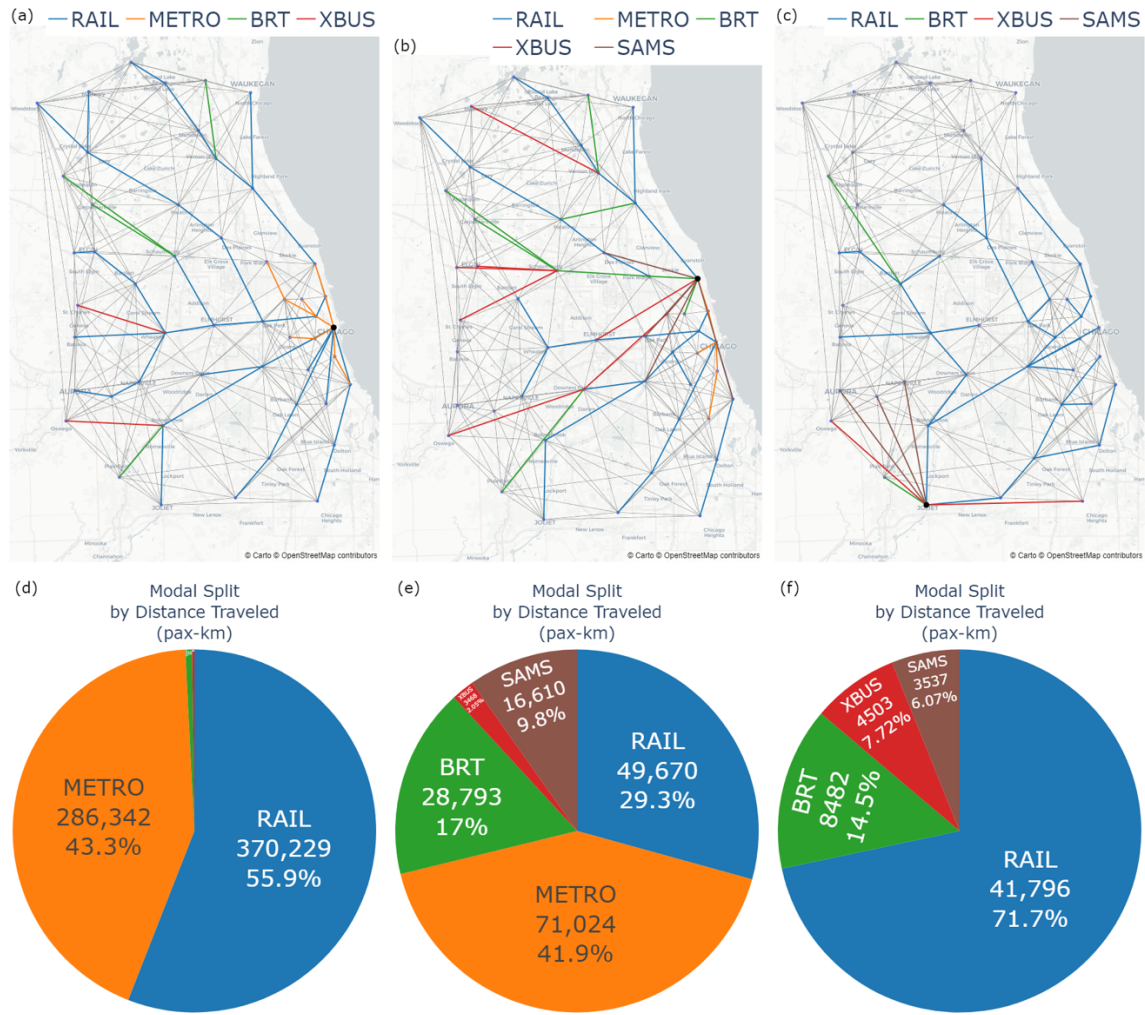

Figure 13. Origin results of Scenario 2 - Existing rail infrastructure plus nominal capital investment (multimodal paths, modal splits by distance traveled) of trips from:
(a,d) Chicago – Loop, (b,e) Chicago – Rogers Park, (c,f) Joliet.

### 5.1.3 Scenario 3 - Greenfield design with no existing rail networks

Scenario 3 redesigns the entire transit network. Figure 14 presents the results that rapid transit rail serves most trips in urban areas and adjacent suburbs, with BRT serving city-suburb and suburb-suburb trips. Compared to the first two scenarios, BRT effectively replaces commuter rail with easier access (smaller stop spacing) and lower capital costs per distance. These allow an extensive BRT network for more direct services and save the budget to extend rapid transit rail to near-city suburbs. Furthermore, interzonal SAMS only cover around 1% of trips, much lower than the previous two scenarios. This shows that multimodal transit networks tailor-made to demand with SAV feeders may resist competition from interzonal SAMS. Compared to Scenario 1, Scenario 3 spends only 76% of equivalent infrastructure cost but saves average journey time by 1.4min/trip (3%) and operating cost by $0.35/trip (7%).

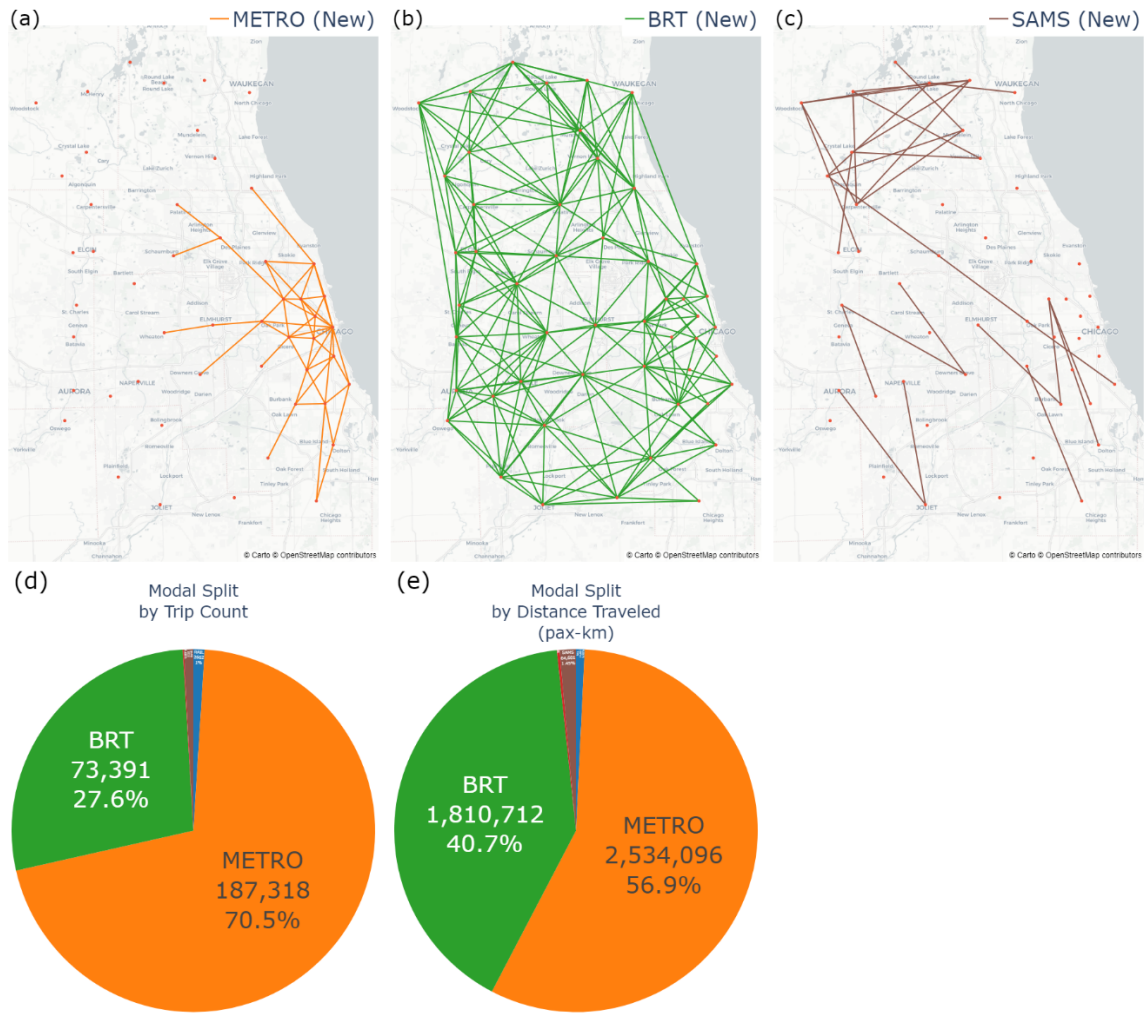

Figure 14. Results of Scenario 3 - Greenfield design with no existing rail infrastructure: (a-c) modal networks (only modes with more than 1% market share are shown), (d) modal split by trip count, (e) modal split by distance traveled.

The specialized roles of *METRO* and *BRT* are further seen in Figure 15 with trips originating from the CBD, city periphery, and suburb. Rapid transit rail covers most trips from the Loop and Rogers Park, with some transfers to BRT required to reach suburban areas further away. For trips from Joliet, BRT is the dominating mode, with some interzonal SAMS reaching areas not covered by BRT.

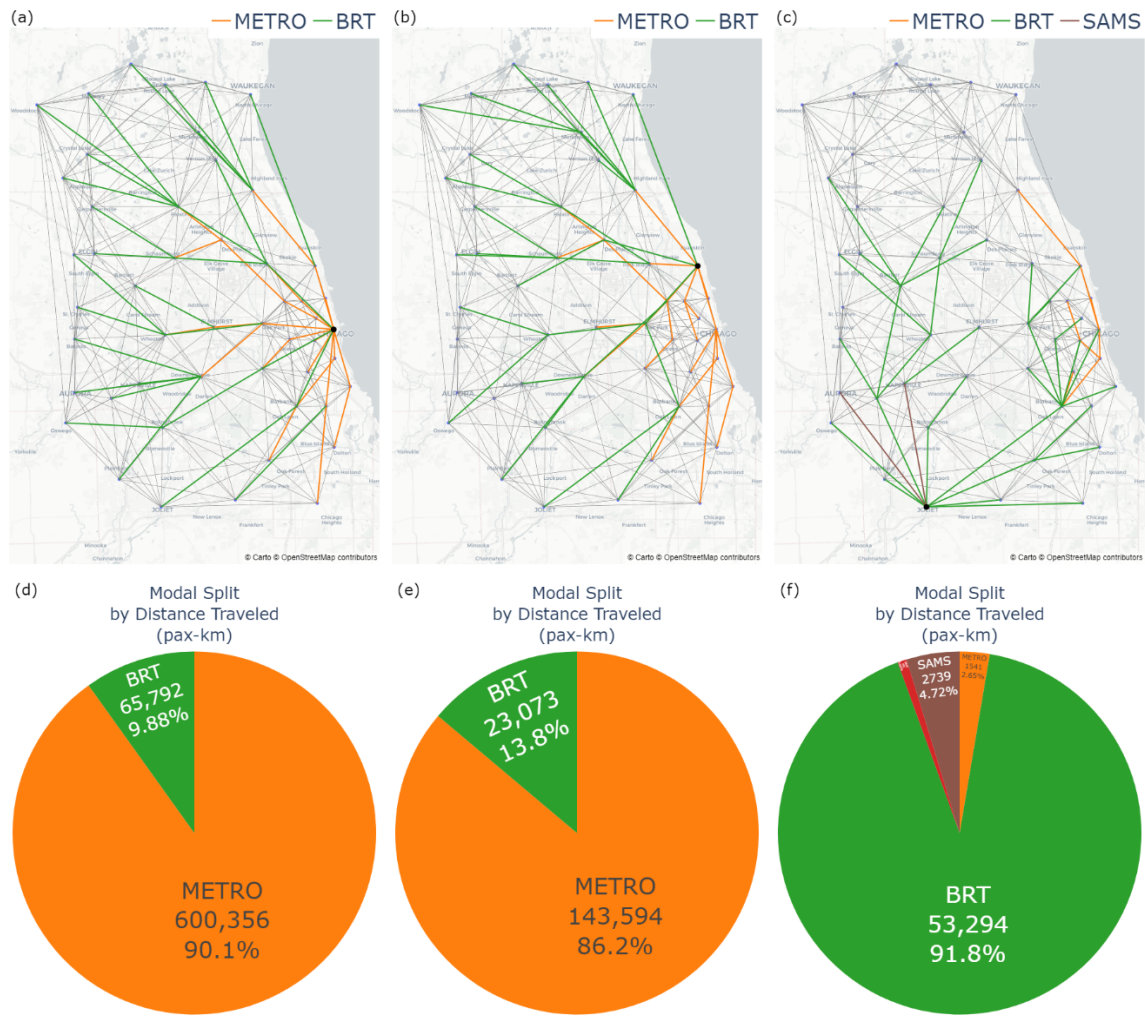

Figure 15. Origin results of Scenario 3 - Greenfield design with no existing rail infrastructure (multimodal paths, modal splits by distance traveled) for trips from:
(a,d) Chicago – Loop, (b,e) Chicago – Rogers Park, (c,f) Joliet.

## 5.2 Route generation

The MILP solver provides exact solutions for route generation within seconds in most cases with the same computational environment in Section 5.1. Only the *RAIL* in Scenarios 1 and 2 took more time (within 30 minutes) to reach a 1% gap, for the more complex connections of multiple routes at some zones. Table 4 compares the computation and scenario results obtained through MILP and myopic heuristics.

The MILP formulation resulted in significant savings of 10% to 71% in intramodal transfers for *RAIL* and *METRO* in Scenarios 1 and 2. These cases, with longer routes and multiple route segments connected in the same zones, pose difficulty for the heuristics to choose the route combinations that can reduce transfers further down the routes. However, in Scenario 3 with the greenfield design, the performance improvement brought by the MILP formulation is limited by the more point-to-point links and fewer intramodal transfers.

Table 4. Routing results with myopic heuristics and MILP formulation.

| *Scenario 1 - Existing rail infrastructure with no new capital investment* | | | |
|---|---|---|---|
| *MILP model characteristics and computation results* | | | |
| **Mode** | **RAIL** | **METRO** | **XBUS** |
| Number of variables: | | | |
| Continuous: | 140,400 | 43,800 | 311,700 |
| Binary: | 1,134 | 354 | 2,664 |
| Number of constraints | 118,521 | 43,597 | 220,021 |
| Computation time (h) | 0.40 | 0.001 | 0.001 |
| Objective | 25,647 | 533 | 2,839 |
| Lower Bound | 25,391 | 533 | 2,839 |
| Solution gap | 1.00% | 0.00% | 0.00% |
| *Scenario results* | | | |
| **Mode** | **RAIL** | **METRO** | **XBUS** |
| Number of intermodal transfers: | | | |
| Myopic heuristics: | 31,873 | 1,850 | 2,918 |
| MILP formulation: | 25,647 | 533 | 2,839 |
| Percentage Difference: | -19.5% | -71.2% | -2.7% |
| Number of routes: | | | |
| Myopic heuristics: | 27 | 7 | 82 |
| MILP formulation: | 20 | 9 | 82 |
| Percentage Difference: | -25.9% | 28.6% | 0.0% |

| *Scenario 2 - Existing rail infrastructure with nominal capital investment* | | | | |
|---|---|---|---|---|
| *MILP model characteristics and computation results* | | | | |
| **Mode** | **RAIL** | **METRO** | **BRT** | **XBUS** |
| Number of variables: | | | | |
| Continuous: | 124,100 | 40,100 | 96,100 | 160,900 |
| Binary: | 986 | 320 | 724 | 1,294 |
| Number of constraints: | 106,046 | 39,925 | 83,347 | 124,616 |
| Computation time (h) | 0.08 | 0.001 | 0.001 | 0.001 |
| Objective | 26,330 | 384 | 907 | 724 |
| Lower Bound | 26,073 | 384 | 907 | 724 |
| Solution gap | 0.99% | 0.00% | 0.00% | 0.00% |
| *Scenario results* | | | | |
| **Mode** | **RAIL** | **METRO** | **BRT** | **XBUS** |
| Number of intermodal transfers: | | | | |
| Myopic heuristics: | 29,124 | 1,228 | 907 | 746 |
| MILP formulation: | 26,330 | 384 | 907 | 724 |
| Percentage Difference: | -9.4% | -68.7% | 0.0% | -3.1% |
| Number of routes: | | | | |
| Myopic heuristics: | 24 | 8 | 35 | 63 |
| MILP formulation: | 22 | 10 | 36 | 64 |
| Percentage Difference: | -8.3% | 25.0% | 2.9% | 1.6% |

### Scenario 3 - Greenfield design with no existing rail infrastructure

*MILP model characteristics and computation results*

| Mode | RAIL | METRO | BRT | XBUS |
|---|---|---|---|---|
| Number of variables: | | | | |
| Continuous: | 79,800 | 186,200 | 536,400 | 10,200 |
| Binary: | 576 | 1,550 | 4,680 | 54 |
| Number of constraints: | 71,984 | 140,335 | 361,110 | 15,811 |
| Computation time (h) | 0.001 | 0.001 | 0.001 | 0.001 |
| Objective | 0 | 23,134 | 12,565 | 0 |
| Lower Bound | 0 | 22,917 | 12,482 | 0 |
| Solution gap | 0.00% | 0.94% | 0.66% | 0.00% |

*Scenario results*

| Mode | RAIL | METRO | BRT | XBUS |
|---|---|---|---|---|
| Number of intermodal transfers: | | | | |
| Myopic heuristics: | 0 | 23,164 | 12,877 | 0 |
| MILP formulation: | 0 | 23,134 | 12,565 | 0 |
| Percentage Difference: | / | -0.1% | -2.4% | / |
| Number of routes: | | | | |
| Myopic heuristics: | 2 | 26 | 118 | 8 |
| MILP formulation: | 2 | 26 | 113 | 8 |
| Percentage Difference: | 0.0% | 0.0% | -4.2% | 0.0% |

*Summary*

| Scenario | 1 | 2 | 3 |
|---|---|---|---|
| Intramodal transfers (MILP formulation of route generation) | 29,019 | 28,400 | 35,699 |
| Intermodal transfers (zonal connection) | 2,447 | 4,915 | 4,417 |
| Average transfers per trip | 0.12 | 0.13 | 0.15 |

Overall, the numbers of transfers (intermodal and intramodal) are low (0.15 or fewer transfers per trip).[19] This supports the previous assumption of relatively small impact of intramodal transfers in zonal connection optimization.

Table 5 offers an illustrative example of the divergence in routing results between the two algorithms using *METRO* in Scenario 1 as a case study. The MILP formulation saves over 70% intramodal transfers compared to the myopic heuristics. A specific example of a superior routing arrangement is extending Route 4 (same route number in both results) from Zone 28 to 32, so one-seat trips are possible from Zone 32 via Zones 28 and 27 to Zones 29 and 31. Conversely, the myopic heuristics can only consider trips between Zones 27, 28, and 32 when determining the routing arrangement at Zone 28. This necessitates more transfers to reach the final destinations.

---

[19] As a reference point, there were 49,433 cross-platform transfers among 317,118 total boardings for the Chicago "L" rapid transit rail system per day in 2022, i.e., around 0.16 transfers per trip. (Chicago Transit Authority, 2023b)

Table 5. Routing results (*METRO*) of myopic heuristics and MILP formulation for Scenario 1 - Existing rail infrastructure with no new capital investment.

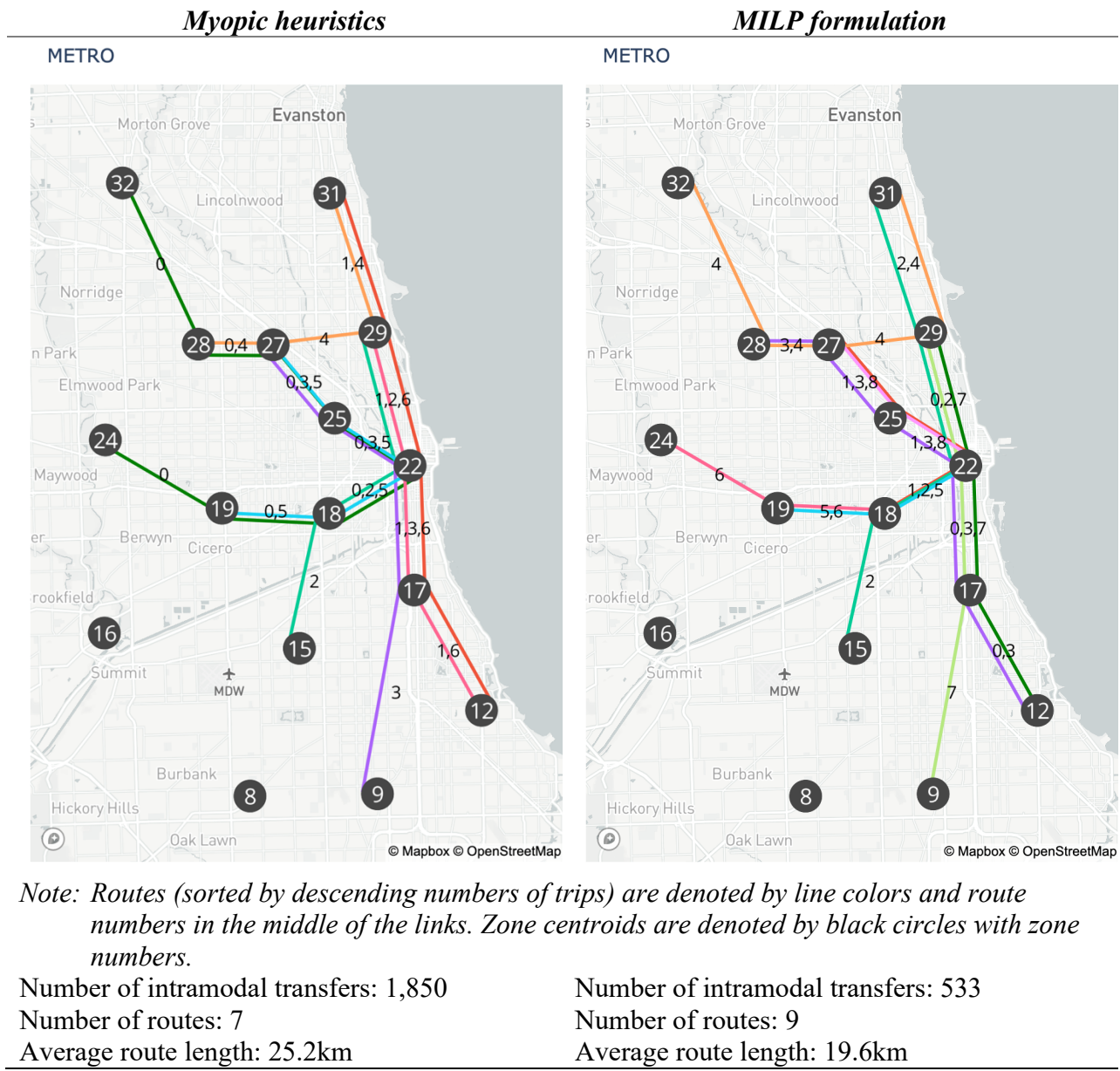

| *Myopic heuristics* | *MILP formulation* |
|---|---|
| Number of intramodal transfers: 1,850 | Number of intramodal transfers: 533 |
| Number of routes: 7 | Number of routes: 9 |
| Average route length: 25.2km | Average route length: 19.6km |

*Note: Routes (sorted by descending numbers of trips) are denoted by line colors and route numbers in the middle of the links. Zone centroids are denoted by black circles with zone numbers.*

# 6 Conclusion

## 6.1 Summary

The growing influence of shared autonomous vehicles (SAVs) presents novel challenges and opportunities for public transit, the backbone of urban mobility. This study develops a methodological framework to redesign a large-scale multimodal transit network, integrating Shared Autonomous Mobility Services (SAMS) as both transit feeders and direct origin-to-destination mode. The proposed method tackles three interconnected sub-problems, namely high-level multimodal modeling, system-optimal link investment and path flows, and route design, addressing the NP-hard transit network design problem (TNDP) and multimodal network interactions.

Our study presents an efficient approach to transit network redesign and route generation that captures the interaction among demand, geospatial, and modal characteristics, and incorporates the existing network. The developed routes are then the input for subsequent joint transit-SAMS service level optimization (frequency and SAV fleet size) and simulation (SAMS fleet and traveler assignment) problem (Pinto et al., 2020).

Conceptually, this study models and solves a transit network design problem simultaneously with *multiple* transit modes and SAMS. The use of continuous approximation to characterize spatial demand and modal performance, along with zonal connection optimization and route generation to minimize total door-to-door generalized costs of users and operators, amplify the study's contributions to the existing literature on transit planning in the era of emerging mobility services.

Theoretically, the zonal connection optimization problem is formulated as an origin-based minimum-cost multi-commodity network flow (MCNF) problem that simultaneously optimizes multimodal links addition and path flows. It is further expanded to the route generation formulation to minimize intramodal transfers.

Methodologically, the origin-based formulation limits the number of variables to represent flows, thereby ensuring efficient solutions but also respects demand and capacity constraints. Strategic consideration of connection candidates also controls the solution space. Three case studies in the Chicago metropolitan area are solved within one to two days, demonstrating the optimization performance and solution quality of the framework. The results highlight the potential of redesigning transit networks with SAV feeders in improving both traveler experience and operator efficiency.

## 6.2 Applications and implications

This study pioneers an efficient, large-scale transit redesign framework that assimilates the emergent SAMS to power transit as feeders or to compete as origin-to-destination services. The resulting synergy between conventional and autonomous mobility solutions offers invaluable insights for planners and policymakers tasked with reimagining transit systems to meet evolving urban mobility demands.

Interzonal SAMS are valuable to cover transit gaps in areas with sparse demand, particularly in infrastructure-limited Scenarios 1 and 2. The computational results favor easily accessible and flexible transit services, such as frequent bus rapid transit (BRT) with closer stops over the faster commuter rail with infrequent services and stops. The comparison of greenfield design in Scenario 3 and the other two scenarios shows that the legacy rail networks may not serve the existing demand well, and newer transit technologies such as BRT may offer a better alternative in new areas. While greenfield design is infeasible in developed cities, these may still provide insights for new development. Overall, multimodal cooperation emerges as an effective strategy for transit agencies to improve service provision and reduce costs, negating the need for long-haul SAV trips.

## 6.3 Limitations and future research

City-scale multimodal transit network design involves solving combinatorial problems, analyzing intricate network interactions, and processing massive demand and geographic datasets. The study presents a foundation framework that enables a system-optimal design and analysis of transit modes with SAMS and allows future refinement and investigation.

First, achieving system-optimal transit networks warrants subsequent decisions on transit frequency, SAV fleet size, and fares to bridge the gap with user equilibrium in terms of mode and route choices. The transit routing results from this study act as precursors to the next-step service level optimization-simulation problem, which also handles user-equilibrium constraints and designs detailed local or intrazonal transit routes and SAMS services. Future works can consider the efficient design and implementation of the iterative loop and synergy between different SAV fleets.

Second, agent-based simulation on top of the redesigned transit network can model demand elasticity not considered in this study, alongside sensitivity to considerations such as intramodal transfers. Improved zonal connectivity and transit services may lead to change in activity schedule and land use, inducing demand change or mode shift from driving. This can evaluate the strengthening of transit economies of density and further improvement following the Mohring effect (1972).

Third, the model does not account for the impact of transit and SAVs on roadway traffic. Given the complexity of road networks and mixed traffic flows between autonomous and human-driven vehicles, a dynamic traffic assignment tool can provide accurate parameters such as link time. Additionally, the dynamic variation of trips in a day can be simulated with multiple model instances of different time horizons.

Fourth, the model would benefit from accurate engineering and geographic inputs. Incorporating cost estimates of infrastructure projects as additional link costs would enable comprehensive cost-and-benefit analyses that capture multimodal interactions with SAMS. Additionally, leveraging Geographic Information Systems (GIS) to obtain precise data on approximating feeder usage and travel time can enhance model accuracy.

Lastly, route generation results could be post-processed for operational considerations with route modification techniques such as short-turning and inter-lining. This provides a diverse set of patterns for the next-step transit frequency optimization.

# 7 Acknowledgements

This paper is based on work that was *partially* supported by the U.S. Department of Energy (DOE) Vehicle Technologies Office (VTO) under the Systems and Modeling for Accelerated Research in Transportation (SMART) Mobility Laboratory Consortium, an initiative of the Energy Efficient Mobility Systems (EEMS) Program. Additional funding was provided by the Northwestern University Transportation Center. The computational experiments were supported in part through the computational resources and staff contributions provided for the Quest high performance computing facility at Northwestern University. The authors remain responsible for all findings and opinions presented in the paper. The contents do not necessarily reflect the views of the sponsoring organizations.

# 8 References

Abdelwahed, A., van den Berg, P.L., Brandt, T., Ketter, W., 2023. Balancing convenience and sustainability in public transport through dynamic transit bus networks. Transportation Research Part C: Emerging Technologies 151, 104100. https://doi.org/10.1016/j.trc.2023.104100

Alonso-Mora, J., Samaranayake, S., Wallar, A., Frazzoli, E., Rus, D., 2017. On-demand high-capacity ride-sharing via dynamic trip-vehicle assignment. PNAS 114, 462–467. https://doi.org/10.1073/pnas.1611675114

American Public Transportation Association, 2020. Bus Rapid Transit Service Design and Operations [WWW Document]. URL https://www.apta.com/wp-content/uploads/APTA-BTS-BRT-RP-004-10_Rev1.pdf (accessed 11.21.23).

Baaj, M.H., Mahmassani, H.S., 1995. Hybrid route generation heuristic algorithm for the design of transit networks. Transportation Research Part C: Emerging Technologies 3, 31–50. https://doi.org/10.1016/0968-090X(94)00011-S

Badia, H., Estrada, M., Robusté, F., 2014. Competitive transit network design in cities with radial street patterns. Transportation Research Part B: Methodological 59, 161–181. https://doi.org/10.1016/j.trb.2013.11.006

Badia, H., Jenelius, E., 2021. Design and operation of feeder systems in the era of automated and electric buses. Transportation Research Part A: Policy and Practice 152, 146–172. https://doi.org/10.1016/j.tra.2021.07.015

Bahbouh, K., Wagner, J.R., Morency, C., Berdier, C., 2017. Travel demand corridors: Modelling approach and relevance in the planning process. Journal of Transport Geography 58, 196–208. https://doi.org/10.1016/j.jtrangeo.2016.12.007

Becker, H., Becker, F., Abe, R., Bekhor, S., Belgiawan, P.F., Compostella, J., Frazzoli, E., Fulton, L.M., Guggisberg Bicudo, D., Murthy Gurumurthy, K., Hensher, D.A., Joubert, J.W., Kockelman, K.M., Kröger, L., Le Vine, S., Malik, J., Marczuk, K., Ashari Nasution, R., Rich, J., Papu Carrone, A., Shen, D., Shiftan, Y., Tirachini, A., Wong, Y.Z., Zhang, M., Bösch, P.M., Axhausen, K.W., 2020. Impact of vehicle automation and electric propulsion on production costs for mobility services worldwide. Transportation Research Part A: Policy and Practice 138, 105–126. https://doi.org/10.1016/j.tra.2020.04.021

Borndörfer, R., Grötschel, M., Pfetsch, M.E., 2007. A Column-Generation Approach to Line Planning in Public Transport. Transportation Science 41, 123–132. https://doi.org/10.1287/trsc.1060.0161

Boutarfa, Z., Gok, M., 2023. Multimodal Public Transit Network Design Method Based on Hub-and-Spoke Infrastructure. Transportation Research Record 03611981231167159. https://doi.org/10.1177/03611981231167159

Calabrò, G., Araldo, A., Oh, S., Seshadri, R., Inturri, G., Ben-Akiva, M., 2023. Adaptive transit design: Optimizing fixed and demand responsive multi-modal transportation via continuous approximation. Transportation Research Part A: Policy and Practice 171, 103643. https://doi.org/10.1016/j.tra.2023.103643

Canca, D., De-Los-Santos, A., Laporte, G., Mesa, J.A., 2019. Integrated Railway Rapid Transit Network Design and Line Planning problem with maximum profit. Transportation Research Part E: Logistics and Transportation Review 127, 1–30. https://doi.org/10.1016/j.tre.2019.04.007

Cancela, H., Mauttone, A., Urquhart, M.E., 2015. Mathematical programming formulations for transit network design. Transportation Research Part B: Methodological 77, 17–37. https://doi.org/10.1016/j.trb.2015.03.006

Ceder, A., Wilson, N.H.M., 1986. Bus network design. Transportation Research Part B: Methodological 20, 331–344. https://doi.org/10.1016/0191-2615(86)90047-0

Chen, H., Gu, W., Cassidy, M.J., Daganzo, C.F., 2015. Optimal transit service atop ring-radial and grid street networks: A continuum approximation design method and comparisons. Transportation Research Part B: Methodological, ISTTT 21 for the year 2015 81, 755–774. https://doi.org/10.1016/j.trb.2015.06.012

Chicago Metropolitan Agency for Planning, 2023. Activity-Based Model (ABM) Calibration and Validation Report.

Chicago Transit Authority, 2023a. Service Standards and Policies [WWW Document]. URL https://www.transitchicago.com/assets/1/6/Chicago_Transit_Authority_Service_Standards.pdf (accessed 11.21.23).

Chicago Transit Authority, 2023b. Annual Ridership Report [WWW Document]. URL https://www.transitchicago.com/assets/1/6/2022_Annual_Report_-_FINAL.pdf (accessed 11.20.23).

Chicago Transit Authority, 2022. CTA - "L" (Rail) Lines - Shapefile [WWW Document]. URL https://data.cityofchicago.org/Transportation/CTA-L-Rail-Lines-Shapefile/53r7-y88m (accessed 6.20.23).

Cipriani, E., Gori, S., Petrelli, M., 2012. Transit network design: A procedure and an application to a large urban area. Transportation Research Part C: Emerging Technologies, Special issue on Optimization in Public Transport+ISTT2011 20, 3–14. https://doi.org/10.1016/j.trc.2010.09.003

City of Chicago, 2012. Metra Lines [WWW Document]. URL https://data.cityofchicago.org/Transportation/Metra-Lines/q8wx-dznq (accessed 6.20.23).

Clarens, G.C., Hurdle, V.F., 1975. An Operating Strategy for a Commuter Bus System. Transportation Science 9, 1–20. https://doi.org/10.1287/trsc.9.1.1

Cortina, M., Chiabaut, N., Leclercq, L., 2023. Fostering synergy between transit and Autonomous Mobility-on-Demand systems: A dynamic modeling approach for the morning commute problem. Transportation Research Part A: Policy and Practice 170, 103638. https://doi.org/10.1016/j.tra.2023.103638

Daganzo, C.F., 2010. Structure of competitive transit networks. Transportation Research Part B: Methodological 44, 434–446. https://doi.org/10.1016/j.trb.2009.11.001

Dandl, F., Engelhardt, R., Hyland, M., Tilg, G., Bogenberger, K., Mahmassani, H.S., 2021. Regulating mobility-on-demand services: Tri-level model and Bayesian optimization solution approach. Transportation Research Part C: Emerging Technologies 125, 103075. https://doi.org/10.1016/j.trc.2021.103075

Durán-Micco, J., Vansteenwegen, P., 2022. A survey on the transit network design and frequency setting problem. Public Transp 14, 155–190. https://doi.org/10.1007/s12469-021-00284-y

Fan, Y., Ding, J., Liu, H., Wang, Y., Long, J., 2022. Large-scale multimodal transportation network models and algorithms-Part I: The combined mode split and traffic assignment problem. Transportation Research Part E: Logistics and Transportation Review 164, 102832. https://doi.org/10.1016/j.tre.2022.102832

Farahani, R.Z., Miandoabchi, E., Szeto, W.Y., Rashidi, H., 2013. A review of urban transportation network design problems. European Journal of Operational Research 229, 281–302. https://doi.org/10.1016/j.ejor.2013.01.001

Federal Transit Administration, 2021. National Transit Summaries & Trends.

Federal Transit Administration, 2017. Greenhouse Gas Emissions from Transit Projects:Programmatic Assessment (No. 0097).

Gong, M., Hu, Y., Chen, Z., Li, X., 2021. Transfer-based customized modular bus system design with passenger-route assignment optimization. Transportation Research Part E: Logistics and Transportation Review 153, 102422. https://doi.org/10.1016/j.tre.2021.102422

Heyken Soares, P., 2021. Zone-based public transport route optimisation in an urban network. Public Transp 13, 197–231. https://doi.org/10.1007/s12469-020-00242-0

Heyken Soares, P., Mumford, C.L., Amponsah, K., Mao, Y., 2019. An adaptive scaled network for public transport route optimisation. Public Transp 11, 379–412. https://doi.org/10.1007/s12469-019-00208-x

Hörl, S., Ruch, C., Becker, F., Frazzoli, E., Axhausen, K.W., 2019. Fleet operational policies for automated mobility: A simulation assessment for Zurich. Transportation Research Part C: Emerging Technologies 102, 20–31. https://doi.org/10.1016/j.trc.2019.02.020

Huang, D., Liu, Z., Fu, X., Blythe, P.T., 2018. Multimodal transit network design in a hub-and-spoke network framework. Transportmetrica A: Transport Science 14, 706–735. https://doi.org/10.1080/23249935.2018.1428234

Hyland, M., Mahmassani, H.S., 2018. Dynamic autonomous vehicle fleet operations: Optimization-based strategies to assign AVs to immediate traveler demand requests. Transportation Research Part C: Emerging Technologies 92, 278–297. https://doi.org/10.1016/j.trc.2018.05.003

Iliopoulou, C., Kepaptsoglou, K., Vlahogianni, E., 2019. Metaheuristics for the transit route network design problem: a review and comparative analysis. Public Transp 11, 487–521. https://doi.org/10.1007/s12469-019-00211-2

Kar, A., Carrel, A.L., Miller, H.J., Le, H.T.K., 2022. Public transit cuts during COVID-19 compound social vulnerability in 22 US cities. Transportation Research Part D: Transport and Environment 110, 103435. https://doi.org/10.1016/j.trd.2022.103435

Kumar, P., Khani, A., 2022. Planning of integrated mobility-on-demand and urban transit networks. Transportation Research Part A: Policy and Practice 166, 499–521. https://doi.org/10.1016/j.tra.2022.11.001

Levin, M.W., Boyles, S.D., 2015. Effects of Autonomous Vehicle Ownership on Trip, Mode, and Route Choice. Transportation Research Record 2493, 29–38. https://doi.org/10.3141/2493-04

Liu, Y., Ouyang, Y., 2021. Mobility service design via joint optimization of transit networks and demand-responsive services. Transportation Research Part B: Methodological 151, 22–41. https://doi.org/10.1016/j.trb.2021.06.005

Luathep, P., Sumalee, A., Lam, W.H.K., Li, Z.-C., Lo, H.K., 2011. Global optimization method for mixed transportation network design problem: A mixed-integer linear programming approach. Transportation Research Part B: Methodological 45, 808–827. https://doi.org/10.1016/j.trb.2011.02.002

Luo, S., Nie, Y. (Marco), 2019. Impact of ride-pooling on the nature of transit network design. Transportation Research Part B: Methodological 129, 175–192. https://doi.org/10.1016/j.trb.2019.09.007

Magnanti, T.L., Wong, R.T., 1984. Network Design and Transportation Planning: Models and Algorithms. Transportation Science 18, 1–55. https://doi.org/10.1287/trsc.18.1.1

Martínez, L.M., Viegas, J.M., Eiró, T., 2015. Formulating a New Express Minibus Service Design Problem as a Clustering Problem. Transportation Science 49, 85–98. https://doi.org/10.1287/trsc.2013.0497

Martínez Mori, J.C., Speranza, M.G., Samaranayake, S., 2023. On the Value of Dynamism in Transit Networks. Transportation Science. https://doi.org/10.1287/trsc.2022.1193

Mo, B., Cao, Z., Zhang, H., Shen, Y., Zhao, J., 2021. Competition between shared autonomous vehicles and public transit: A case study in Singapore. Transportation Research Part C: Emerging Technologies 127, 103058. https://doi.org/10.1016/j.trc.2021.103058

Mohring, H., 1972. Optimization and Scale Economies in Urban Bus Transportation. The American Economic Review 62, 591–604.

Narayanan, S., Chaniotakis, E., Antoniou, C., 2020. Shared autonomous vehicle services: A comprehensive review. Transportation Research Part C: Emerging Technologies 111, 255–293. https://doi.org/10.1016/j.trc.2019.12.008

Newell, G.F., 1979. Some Issues Relating to the Optimal Design of Bus Routes. Transportation Science 13, 20–35. https://doi.org/10.1287/trsc.13.1.20

Ng, M.T.M., Mahmassani, H.S., 2023. Autonomous Minibus Service With Semi-on-Demand Routes in Grid Networks. Transportation Research Record 2677, 178–200. https://doi.org/10.1177/03611981221098660

Pinto, H.K.R.F., Hyland, M.F., Mahmassani, H.S., Verbas, I.Ö., 2020. Joint design of multimodal transit networks and shared autonomous mobility fleets. Transportation Research Part C: Emerging Technologies, ISTTT 23 TR_C-23rd International Symposium on Transportation and Traffic Theory (ISTTT 23) 113, 2–20. https://doi.org/10.1016/j.trc.2019.06.010

Rennert, K., Errickson, F., Prest, B.C., Rennels, L., Newell, R.G., Pizer, W., Kingdon, C., Wingenroth, J., Cooke, R., Parthum, B., Smith, D., Cromar, K., Diaz, D., Moore, F.C., Müller, U.K., Plevin, R.J., Raftery, A.E., Ševčíková, H., Sheets, H., Stock, J.H., Tan, T., Watson, M., Wong, T.E., Anthoff, D., 2022. Comprehensive evidence implies a higher social cost of $CO_2$. Nature 610, 687–692. https://doi.org/10.1038/s41586-022-05224-9

Ritchie, H., 2020. Cars, planes, trains: where do $CO_2$ emissions from transport come from? [WWW Document]. Our World in Data. URL https://ourworldindata.org/co2-emissions-from-transport (accessed 10.14.22).

Salazar, M., Lanzetti, N., Rossi, F., Schiffer, M., Pavone, M., 2020. Intermodal Autonomous Mobility-on-Demand. IEEE Transactions on Intelligent Transportation Systems 21, 3946–3960. https://doi.org/10.1109/TITS.2019.2950720

Shan, A., Hoang, N.H., An, K., Vu, H.L., 2021. A framework for railway transit network design with first-mile shared autonomous vehicles. Transportation Research Part C: Emerging Technologies 130, 103223. https://doi.org/10.1016/j.trc.2021.103223

Shen, Y., Zhang, H., Zhao, J., 2018. Integrating shared autonomous vehicle in public transportation system: A supply-side simulation of the first-mile service in Singapore. Transportation Research Part A: Policy and Practice 113, 125–136. https://doi.org/10.1016/j.tra.2018.04.004

Sieber, L., Ruch, C., Hörl, S., Axhausen, K.W., Frazzoli, E., 2020. Improved public transportation in rural areas with self-driving cars: A study on the operation of Swiss train lines. Transportation Research Part A: Policy and Practice 134, 35–51. https://doi.org/10.1016/j.tra.2020.01.020

Sivakumaran, K., Li, Y., Cassidy, M., Madanat, S., 2014. Access and the choice of transit technology. Transportation Research Part A: Policy and Practice 59, 204–221. https://doi.org/10.1016/j.tra.2013.09.006

Steiner, K., Irnich, S., 2020. Strategic Planning for Integrated Mobility-on-Demand and Urban Public Bus Networks. Transportation Science 54, 1616–1639. https://doi.org/10.1287/trsc.2020.0987

Tahlyan, D., Said, M., Mahmassani, H., Stathopoulos, A., Walker, J., Shaheen, S., 2022. For whom did telework not work during the Pandemic? understanding the factors impacting telework satisfaction in the US using a multiple indicator multiple cause (MIMIC) model.

Transportation Research Part A: Policy and Practice 155, 387–402. https://doi.org/10.1016/j.tra.2021.11.025

The Institute of Transportation and Development Policy (ITDP), 2018. The Bus Rapid Transit Planning Guide, 4th ed.

U.S. Bureau of Labor Statistics, 2023. Producer Price Index by Industry: Transportation Industries [WWW Document]. FRED, Federal Reserve Bank of St. Louis. URL https://fred.stlouisfed.org/series/PCUATRANSATRANS (accessed 6.15.23).

US Department of Transportation, 2016. The Value of Travel Time Savings: Departmental Guidance for Conducting Economic Evaluations Revision 2 (2016 Update).

Verbas, İ.Ö., Mahmassani, H.S., Hyland, M.F., 2015. Dynamic Assignment-Simulation Methodology for Multimodal Urban Transit Networks. Transportation Research Record 2498, 64–74. https://doi.org/10.3141/2498-08

Viggiano, C., Koutsopoulos, H.N., Wilson, N.H.M., Attanucci, J., 2018. Applying Spatial Aggregation Methods to Identify Opportunities for New Bus Services in London. Transportation Research Record 2672, 75–85. https://doi.org/10.1177/0361198118797218

Wang, Yineng, Lin, X., He, F., Li, M., 2022. Designing transit-oriented multi-modal transportation systems considering travelers' choices. Transportation Research Part B: Methodological 162, 292–327. https://doi.org/10.1016/j.trb.2022.06.002

Wang, Yu, Liu, H., Fan, Y., Ding, J., Long, J., 2022. Large-scale multimodal transportation network models and algorithms-Part II: Network capacity and network design problem. Transportation Research Part E: Logistics and Transportation Review 167, 102918. https://doi.org/10.1016/j.tre.2022.102918

Wardman, M., 2004. Public transport values of time. Transport Policy 11, 363–377. https://doi.org/10.1016/j.tranpol.2004.05.001

Xu, X., Mahmassani, H.S., Chen, Y., 2019. Privately owned autonomous vehicle optimization model development and integration with activity-based modeling and dynamic traffic assignment framework. Transportation Research Record 2673, 683–695. https://doi.org/doi.org/10.1177/0361198119852072

Ye, J., Jiang, Y., Chen, J., Liu, Z., Guo, R., 2021. Joint optimisation of transfer location and capacity for a capacitated multimodal transport network with elastic demand: a bi-level programming model and paradoxes. Transportation Research Part E: Logistics and Transportation Review 156, 102540. https://doi.org/10.1016/j.tre.2021.102540

Yuan, Y., Yu, J., 2018. Locating transit hubs in a multi-modal transportation network: A cluster-based optimization approach. Transportation Research Part E: Logistics and Transportation Review 114, 85–103. https://doi.org/10.1016/j.tre.2018.03.008

Zhang, L., Yang, H., Wu, D., Wang, D., 2014. Solving a discrete multimodal transportation network design problem. Transportation Research Part C: Emerging Technologies 49, 73–86. https://doi.org/10.1016/j.trc.2014.10.008

# Appendix A  Zonal connection optimization - minimum flow constraint

Following the formulation for zonal connection optimization in Section 3.3, it is possible to set a lower bound for the transit link flows based on the minimum service standard with a maximum policy headway to ensure the integrality of vehicles. This aligns with the assumed vehicle occupancy and inferred waiting time from headway, and also allows modeling of abandoned service for existing rail services.

Eq. (A-1) provides the lower bound of the transit link flow given the number of links $x^\lambda_{mij}$, with minimum policy occupancy $R^{min}_m$ and maximum policy headway $H^{max}_m$. This ensures the assigned flow is sufficient for the economies of density assumed in the operating costs, while headway is not too high to violate the assumed average waiting time. However, the design occupancy and headway are not used, as the costs can be counterbalanced by some dense links.

$$\sum_{o \in \mathcal{Z}} \frac{f^\lambda_{omij} + f^\lambda_{omji}}{2} \geq \frac{R^{min}_m}{H^{max}_m} x^\lambda_{mij}, \forall (i,j) \in \mathcal{L}_m, \forall m \in \mathcal{M}^t \tag{A-1}$$

While the MILP model minimizes total generalized costs subject to these additional lower-bound constraints, it may create superfluous "backflows" to satisfy the minimum flows needed to activate a link for a mode. To avoid backflows, Eq. (A-2) introduces binary directional flow variables $x^\varphi_{oij}$ which equals 1 only when the flow originating at zone $o \in \mathcal{Z}$ flows from zone $i \in \mathcal{Z}$ to $j \in \mathcal{Z}$, and 0 otherwise. Eq. (A-3) ensures uni-directional flows from an origin for all links, while Eq. (A-4) is a Big-M constraint setting an upper bound and therefore directions for the flow accordingly, where $M$ is a sufficiently large constant.

$$x^\varphi_{oij} \in \{0,1\}, \forall o \in \mathcal{Z}, (i,j) \in \mathcal{L}_m, \forall m \in \mathcal{M} \tag{A-2}$$

$$x^\varphi_{oij} + x^\varphi_{oji} \leq 1, \forall o \in \mathcal{Z}, (i,j) \in \mathcal{L}_m, \forall m \in \mathcal{M} \tag{A-3}$$

$$f^\lambda_{omij} \leq M x^\varphi_{oij}, \forall o \in \mathcal{Z}, (i,j) \in \mathcal{L}_m, \forall m \in \mathcal{M} \tag{A-4}$$

Nevertheless, these constraints may be unnecessary as it is possible to provide lower service levels with interlining or shorter trains (for rail) without excessive operating costs or perceived waiting costs. The waiting time is less of a concern for peak-hour scheduled services, as commuters time their journey with transit services. The additional constraints also pose additional computational burdens. Furthermore, backflows are only observed in limited cases and scales. Therefore, it seems more practical to first add Eq. (A-1) after observing prevalent small flows in some links, and variables and constraints in Eq. (A-2)-(A-4) after observing backflows.

# Appendix B  Computational results – modal split by distance

This appendix supplements the result discussion in Section 5.1 by exhibiting the mode splits by trip distance in the three scenarios. It is reminded that the results only account for the *interzonal* connections but not the intrazonal trips.

Figure B-1 shows the split by distance traveled of four modes used in Scenario 1. Commuter rail is dominant in long-distance travel, contributing to more than half in trips above 20km. In contrast, rapid transit rail primarily serves the shorter trips, predominately over 70% among those within 10km. SAMS is mostly used for middle-distance trips and express buses for longer trips.

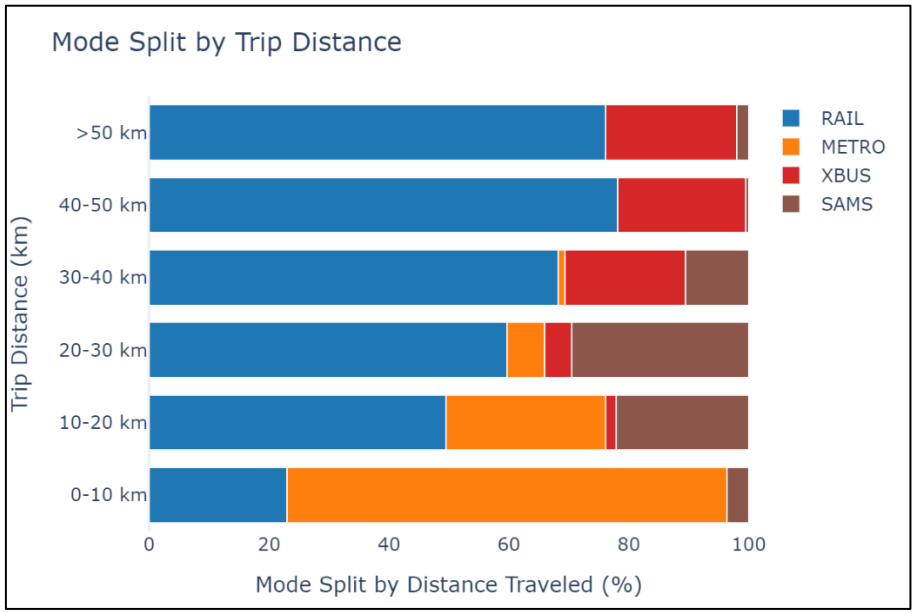

Figure B-1. Mode Split Categorized by Trip Distance for
Scenario 1 – Existing rail infrastructure with no infrastructure investment.

For Scenario 2 shown in Figure B-2, the results are largely similar to Scenario 1 except for the introduction of BRT. BRT mainly replaces some metro rail and SAMS for shorter trips below 20km, and express buses at longer trips. This is related to the designed network which covers limited city area and some suburbs.

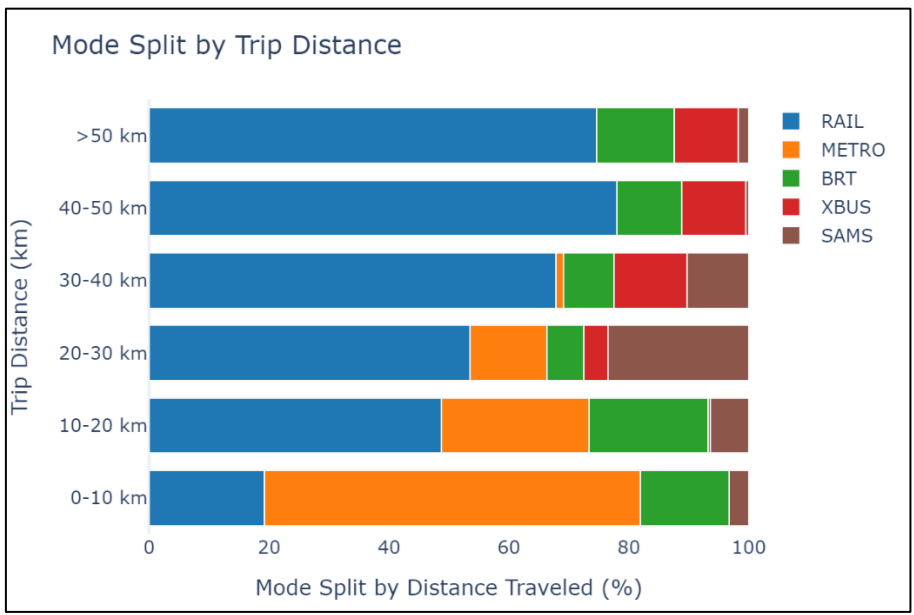

Figure B-2. Mode Split Categorized by Trip Distance for
Scenario 2 - Existing rail infrastructure with nominal infrastructure investment.

For a greenfield design in Figure B-3, Scenario 3 mostly relies on metro rail for the city and city-to-near-suburb trips and BRT for long distance trips. SAMS still persist in limited middle distance trips (20-40km).

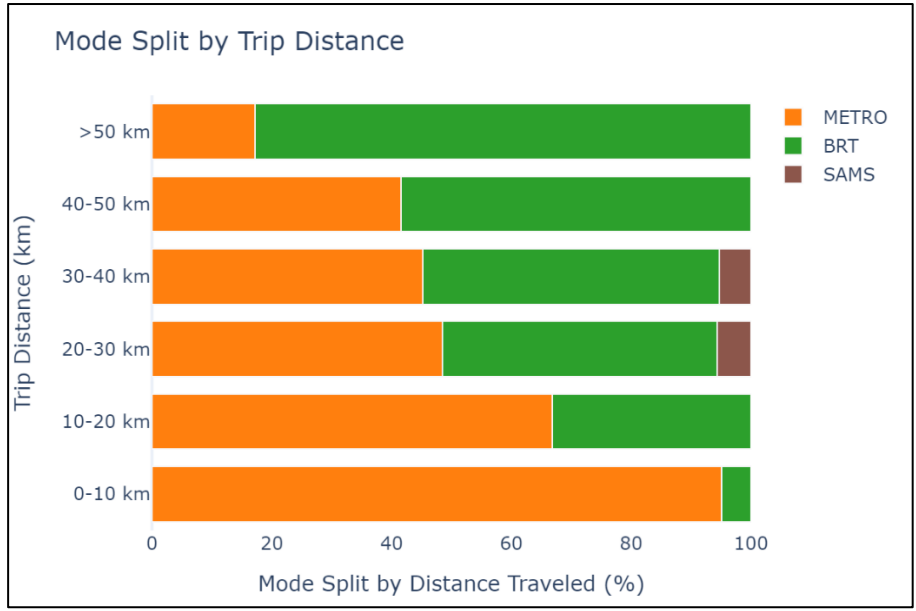

Figure B-3. Mode Split Categorized by Trip Distance for Scenario 3 - Greenfield design with no existing rail networks *(only modes with more than 1% market share are shown)*.